\documentclass[namedreferences]{solarphysics}

\usepackage[hyperref,optionalrh]{spr-sola-addons} 
\usepackage{graphicx}        
\usepackage{color}           
\usepackage{float}			
\usepackage{textcomp}

\usepackage{gensymb}		
\usepackage{epstopdf}
\newcommand{\corr}[1]{{#1}}

\chardef\us=`\_

\begin{document}

\begin{article}
\begin{opening}

\title{Understanding the Role of Mass-Unloading in a Filament Eruption}

\author[addressref=aff1,corref,email={jack.jenkins.16@ucl.ac.uk}]{\inits{J.M.}\fnm{J.~M.}~\lnm{Jenkins}\orcid{0000-0002-8975-812X}}
\author[addressref=aff1]{\inits{D.M.}\fnm{D.~M.}~\lnm{Long}\orcid{0000-0003-3137-0277}}
\author[addressref={aff1,aff2,aff3}]{\inits{L.V.D.}\fnm{L.}~\lnm{van~Driel-Gesztelyi}\orcid{0000-0002-2943-5978}}
\author[addressref=aff4]{\inits{J.C.}\fnm{J.}~\lnm{Carlyle}\orcid{0000-0001-9633-6091}}

\address[id=aff1]{University College London, Mullard Space Science Laboratory, Holmbury St. Mary, Dorking, Surrey, RH5 6NT, UK}
\address[id=aff2]{LESIA-Observatoire de Paris, CNRS, UPMC Univ Paris 06, Univ. Paris-Diderot, F-92195, Meudon Cedex, France}
\address[id=aff3]{Konkoly Observatory of the Hungarian Academy of Sciences, Budapest, Hungary}
\address[id=aff4]{European Space Agency, ESTEC, Noordwijk, Netherlands}

\runningauthor{Jenkins et al.}
\runningtitle{Understanding the Role of Mass-Unloading in a Filament Eruption}

\begin{abstract}

	We describe a partial filament eruption on 11 December 2011 which demonstrates that the inclusion of mass is an important next step for understanding solar eruptions. Observations from the \textit{Solar Terrestrial Relations Observatory Behind} (STEREO-B) and the \textit{Solar Dynamics Observatory} (SDO) spacecraft were used to remove line-of-sight projection effects in filament motion and correlate the effect of plasma dynamics with the evolution of the filament height. Flux cancellation and nearby flux emergence are shown to have played a role in increasing the height of the filament prior to eruption. The two viewpoints allow the quantitative estimation of a large mass-unloading, the subsequent radial expansion, and the eruption of the filament to be investigated. A 1.8 to 4.1 lower-limit ratio between gravitational and magnetic tension forces was found. We therefore conclude that following the loss-of-equilibrium of the flux rope, the radial expansion of the flux rope was restrained by the filamentary material until 70\% of the mass had evacuated the structure through mass-unloading.
\end{abstract}
\keywords{Coronal Mass Ejections, initiation and propagation $\bullet$ Prominence, quiescent $\bullet$ Magnetic fields, photosphere}
\end{opening}

\section{Introduction}
     \label{S-Introduction}

	Multi-wavelength observations of the solar atmosphere reveal different features and phenomena depending on the emission/absorption characteristics of the material that is being observed. Frequently studied large-scale features include active regions, flares, filaments and prominences, and coronal mass ejections (CMEs). Eruptive activity on the Sun is largely associated with flares and CMEs, believed to be triggered by non-ideal and ideal instabilities respectively. For a more detailed summary of solar features and their eruptions, see reviews by \citet{Forbes:2000}, \citet{Webb:2012}, \citet{Parenti:2014}, and references therein.

	Filaments and prominences show a minor preference to form with an average latitude of $\pm$25$\degree$ due to the presence of the highly-sheared field across the polarity inversion line (PIL) of active regions, but they are also seen to exist at higher latitudes \citep[][]{Mackay:2008, McIntosh:2014}. Their difference in appearance is attributed to being different projections due to the parameters employed in observations, but are fundamentally the same phenomena. Filaments are observationally identified as dynamic, dark, and elongated slab-like features against the disc and are thus seen in absorption. Observations of filaments require that the wavelengths used correspond to that which is either absorbed or volume-blocked (reflected) by the material present in the filament. Their off-limb counterparts, prominences, exhibit both similar and different plane-of-sky dynamics and features to filaments \citep[][]{Gunar:2015}. Prominences are also observed as elongated structures but are seen in emission protruding from the surface and appearing above the limb, requiring that the wavelengths used in observations correspond to the emission of the material contained within the prominence. Early spectroscopic observations of the solar atmosphere (\textit{i.e.,} above the photosphere) revealed the presence of highly ionised material such as Fe~{\sc xiv} (previously thought to be the new element, coronium), requiring effective temperatures of $>$~10$^{6}$~K \citep[][]{Edlen:1943}. Observations of off-limb prominences reveal that the observed material is in fact best seen in the H$\alpha$~6562.8~\AA\ and He~{\sc ii}~304~\AA\ passbands, requiring a much cooler effective plasma temperature of 10$^{3}$~-~10$^{4}$~K, indicative of chromospheric material. The consensus is that filaments/prominences consist of this chromospheric plasma suspended in, and thermally isolated from, the hotter corona. For a more in-depth review of filament and prominence observations, see \citet{Parenti:2014}.
        
	The suspension of this chromospheric plasma is potentially facilitated by a system of helical field lines. \citet{vanballegooijen:1989} describe a model based on a dynamically formed system gradually becoming disconnected from the surface at points of local reconnection. This helical system is referred to as a flux rope and forms above a PIL. The evolution of this magnetically buoyant structure then depends on the relationship between magnetic tension, magnetic pressure gradient, and, if the structure contains plasma, gravity.  
    
    The magnetic models developed to study this loss-of-balance of forces internal to the flux rope are mainly based around the two ideal magnetohydrodynamic (MHD) instabilities, the kink and torus instabilities \citep[\textit{e.g.,}][]{Torok:2005, Kliem:2006}. The kink instability is thought to occur when the amount of internal twist in a flux rope, \textit{i.e.,} the number of turns in the field lines around the axis of the flux rope, exceeds some critical value. Such a highly wound flux rope then evolves to reduce this strong internal twist by transferring some of its twist into writhe (twist of the flux rope axis), conserving helicity in the process \citep{Hood:1981}. This increase in writhe includes an associated increase in the height and curvature of the apex of the flux rope, introducing a radial gradient in the magnetic pressure between the regions below and above the flux rope. If this gradient is sufficiently large \citep[historically indicated by the value of the decay index $n_c$~$>$~$3/2$,][]{Bateman:1978}, the pressure gradient (hoop force) drives an exponential rise of the flux rope. The triggering of this exponential expansion due to the large gradient in the magnetic pressure force is referred to as the onset of the torus instability.

	In addition to the internal evolution of the magnetic field of a flux rope, the interaction with external magnetic structures and dynamics have been studied to probe their role in the evolution of a filament channel. \citet{Feynman:1995} and \citet{Chen:2000} describe how the orientation of co-located flux emergence can have different effects on the stability of a nearby filament. Flux emergence manifests itself in the photospheric line-of-sight (LOS) magnetic field data as opposite polarities that emerge, grow, and separate. The observational marker of emergence in the optical--extreme ultraviolet (EUV) wavelengths is often dynamic dark loops forming low in the chromosphere and expanding into the corona, whilst being heated and becoming bright. Although the initial emergence of flux typically contains the observational markers mentioned, the amount of flux that emerges and evolves can vary in scale. A more detailed description on the physics of flux emergence may be found in \citet[][]{Cheung:2014}. The flux emergence associated with filament eruptions can either be located directly beneath the flux rope, as in \citet[][]{Palacios:2015}, or to the side. Emergence beneath the flux rope can trigger the eruption through tether-cutting as in \citet[][]{Moore:2001}. Alternatively, emergence to the side of the flux rope can cause reconnection between the two systems to occur, such that the eruption is triggered by the weakening of the overlying field tension \citep[][]{Ding:2008}. Both of these scenarios are believed to be eruption triggers whilst the eruption drivers are believed to still be instabilities such as the torus instability.

	The initial studies of the formation, stability, evolution, and eruption of filaments led the solar community to the conclusion that the evolution of the magnetic environment containing the filament was of primary importance. \citet{Demoulin:1998} described the need for intrinsic concave-up portions of the magnetic field structure to be able to maintain the suspension of the plasma of a filament, as modelled in \citet{vanballegooijen:1989}, and simulated by authors such as \citet{Lionello:2002} and \citet{Aulanier:2010}. This configuration can presumably be formed on the order of days or weeks and was thus in line with the time-line of filament formation according to observations. As a result, recent numerical models describing the destabilisation of flux ropes have focused mainly on the magnetic evolution, neglecting plasma processes associated with the filament contained within the flux rope. However, \citet{Demoulin:1998} also concluded that the plasma processes are capable of modifying the magnetic environment on timescales of an hour. Similarly, \citet[][]{Gunar:2013} examined non-linear force-free magnetic dip models, concluding that the exclusion/inclusion of plasma could dramatically affect the \corr{radial} B$_{z}$ component of the dips within the magnetic field of a flux rope. This deformation of field lines through the addition or removal of plasma within the magnetic environment on short timescales is a key part of the mass-loading eruption mechanism described in \citet{Low:1996} and \citet{Klimchuk:2001}. The mass-loading model suggests that a sufficiently large mass, such as the filamentary plasma, could allow the flux rope to build free magnetic energy without losing equilibrium until this mass is removed. Once the equilibrium is lost due to the unloading of the anchoring force supplied by the filamentary material, the magnetic environment would be free to expand \citep{Forbes:2000} and attempt to find a new equilibrium by increasing the height of the flux rope, however this does not necessarily result in an eruption.

	Unfortunately, the observations required to accurately correlate the effect of the magnetic environment and internal/external dynamics on the overall height--time (h--t) evolution of the flux rope are rare. However, the ideal conditions for drawing connections between filament mass dynamics and their relation to the h--t profile were demonstrated by \citet{Seaton:2011}. Their analysis of the event on 3 April 2010 describes an active region filament observed from multiple perspectives. Observations from the \textit{Atmospheric Imaging Assembly} \citep[AIA;][]{Lemen:2012} on board the \textit{Solar Dynamics Observatory} \citep[SDO;][]{Pesnell:2012} and the \textit{Extreme Ultraviolet Imager} \citep[EUVI;][]{Wuelser:2004} on board \textit{Solar Terrestrial Relations Observatory Behind} \citep[STEREO;][]{Kaiser:2008} isolated the h--t response to mass-flow dynamics and captured one of the first well documented examples of the mass-loading eruption mechanism. Their event displayed mass motion towards the footpoints of the filament of interest prior to eruption. This unloading of filament mass was then seen to precede the increase in height of the filament spine until catastrophic loss-of-equilibrium set in and the eruption occurred. However \citet[][]{Seaton:2011} did not quantify the degree of mass-unloading undergone in this event, nor did they quantitatively discuss this effect on the evolution of the entire filament.%
    
	Work completed by \citet{Bi:2014} on the filament eruption of 23 February 2012 describes a filament eruption also seen from multiple perspectives. Using the opacity method and application outlined in \citet{Williams:2013} and \citet{Carlyle:2014a} respectively, the authors were able to estimate the amount of mass contained within the filament structure prior to eruption, and the degree of mass-unloading undergone with respect to the entire filament mass. Quantifying the degree of mass-unloading gives further insight into the physics at play but no conclusions on the energetics of the event are presented. Similarly the exact response of the h--t profile of the filament to the described plasma processes during the pre-eruption phase \corr{is omitted}.
 
	In this manuscript, we discuss how the plasma within and magnetic field around a filament evolved prior to its partial eruption on 11 December. The event was observed by multiple observatories located at different points in the heliosphere, providing an opportunity to fully describe the evolution of the filament plasma. In Section~\ref{S-roadmap} we present the outline of the paper. In Section~\ref{S-observations} we describe the observations made, before outlining the results of analysing \corr{these data} in Section~\ref{S-results}. The results are then interpreted and described in Section~\ref{S-discussion}, before some conclusions are drawn in Section~\ref{S-conclusions}. %

%
     

\section{Overview} 
  \label{S-roadmap}      

\begin{figure}
 \centerline{
 			\includegraphics[width=1\textwidth,clip=, trim=20 0 30 0]{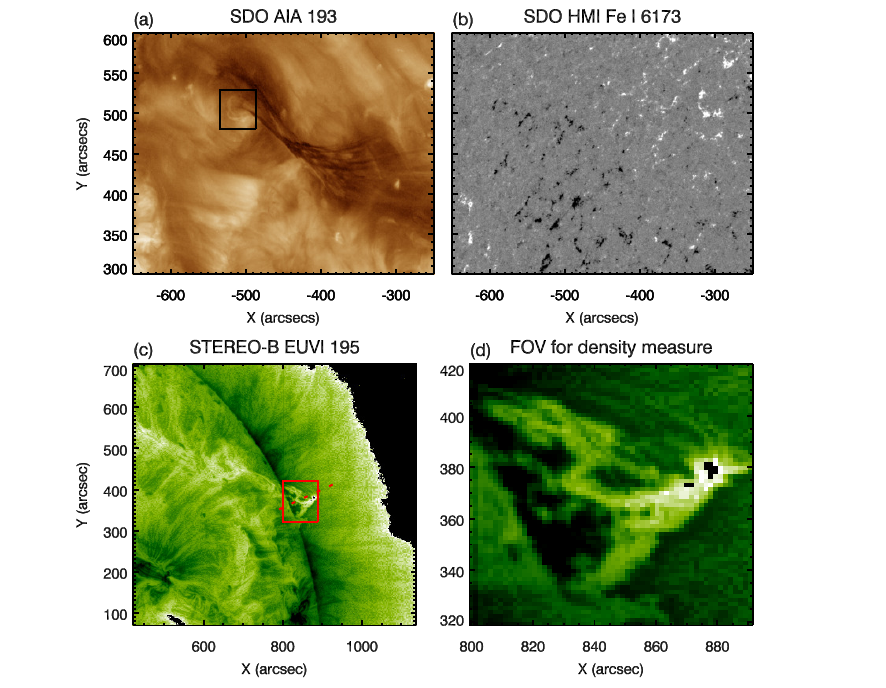}
            }
 \caption{The filament and its photospheric magnetic field environment as seen from the perspective of STEREO-B and SDO. \textit{Panel a}; The filament as seen in the SDO/AIA 193~\AA\ passband. Black box represents the field-of-view (FOV) used for the density measures of Figure~\ref{fig:sdo_density_motion}. \textit{Panel b}; The same FOV as in \textit{(a)} from the SDO/HMI instrument showing the line-of-sight (LOS) magnetic field saturated to $\pm$~100~G. \textit{Panel c}; The filament (indicated by the red box) as seen on the limb by the STEREO-B/EUVI 195~\AA\ passband using a reversed color table. The red dashed line indicates location of the stack line for Figure~\ref{fig:kinematics}. Red box represents the zoomed in FOV shown in \textit{panel d} and used for the density measures of Figure~\ref{fig:stereo_mass_density}. All EUV images have the time stamp of 04:51~UT on 11 December 2011 and the STEREO-B/EUVI images have been processed using the multi-Gaussian normalisation technique \citep[MGN;][]{Morgan:2014}.}\label{fig:context}%
\end{figure}

In an effort to make the key events described in this paper easier to follow, we will now provide a brief summary of the observations made. The partial filament eruption studied here was observed on 11 December 2011 in the north-eastern quadrant of the solar disk by SDO and at the north-western limb by STEREO-B, as shown in Figure~\ref{fig:context}. The analysis described here focuses on the period leading up to the eruption, specifically from 12:00~UT on 10 December to 08:00~UT on 11 December.

\begin{figure}
 \centerline{\hspace*{0.005\textwidth}
 			\includegraphics[width=1\textwidth,clip=, trim=0 0 0 0]{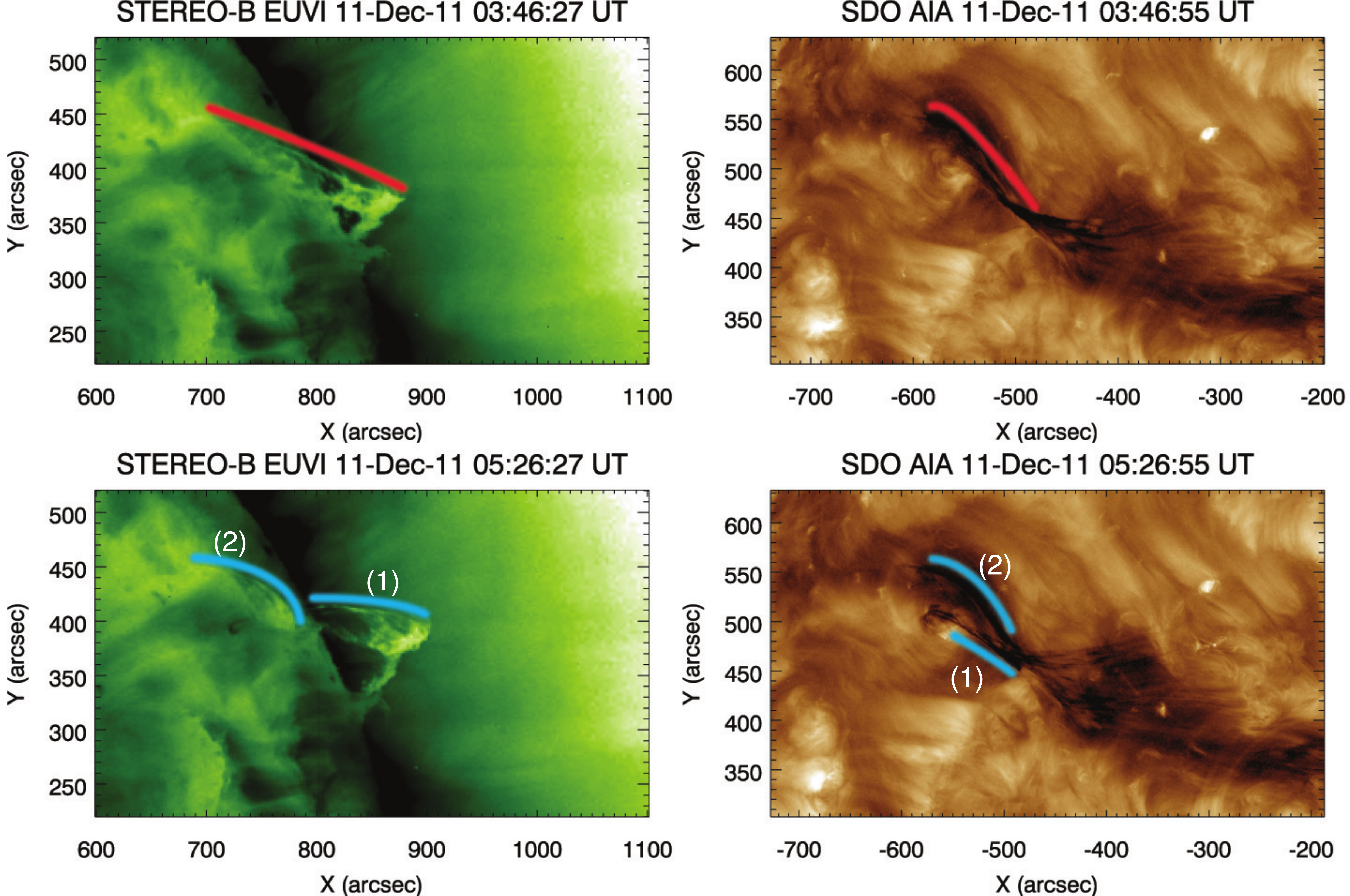}
            }\vspace*{0.01\textwidth}
 \caption{The splitting of the filament during its eruption as seen by STEREO-B (left, reverse colour table) and SDO (right). \textit{Upper}; The red line traces the connection between the upper and lower branches of the filament (STEREO-B) and the south-western and north-eastern portions of the filament (SDO), indicating a magnetic connection highlighted by the illuminating material. \textit{Lower}; The blue lines trace the edges of the two distinctly separated filament structures, (1)--dynamic portion (2)--restrained portion, just prior to the eruption of the dynamic portion of the filament. A movie of this figure accompanies the online version of this manuscript.}\label{fig:splitting}%
\end{figure}

\begin{enumerate}
\item The filament of interest was one of several located in a large filament channel that spanned approximately half of the solar disk visible from SDO/AIA. LOS magnetic field observations from the HMI show that the filament channel was flanked by a very diffuse bipolar photospheric field, common for quiescent filament channels \citep[][]{Mackay:2008}. Approximately 18 hours before the eruption, flux cancellation was recorded along the PIL of this weak bipolar field. During this time, observations from STEREO-B/EUVI showed the filament of interest increasing in height. The flux cancellation along the PIL was then seen to have ceased approximately 12 hours before the eruption. This is discussed in more detail in Section~\ref{S-flux}.

\item Approximately nine hours prior to the eruption, a small bipole was observed to emerge to the north-west of the filament. The orientation of the bipole was perpendicular to the axis of the filament channel. This small bipole grew in extent, reaching its peak value approximately five hours before the eruption of the filament and subsequently began to decay. As the bipole approached its peak strength the filament ceased rising, remaining stationary for approximately one hour. This is discussed in Section~\ref{S-flux-emergence}.

\item The filament was then seen to become unstable, potentially due to the associated flux rope becoming kink unstable, and began a shallow exponential expansion through the corona. Observations from STEREO-B/EUVI suggest that the rising filament did not remain parallel to the surface, see Figure~\ref{fig:splitting}. Shortly after the expansion of the filament restarted, mass was observed flowing from the apex of the filament towards the north--eastern footpoint as observed by SDO. When the filament apex reached a height of approximately 65--70~Mm, approximately one and a half hours prior to the eruption, a large mass flow was observed draining from the apex down to the north--eastern footpoint of the filament. Shortly after the initiation of the large mass flow the expansion of the filament dramatically accelerated. This is discussed in detail in Sections~\ref{S-flows}, \ref{S-density}, and \ref{S-kinematics}.
\end{enumerate}
During the eruption the filament is seen to split in two, shown in Figure~\ref{fig:splitting}, suggesting that the magnetic structure containing the filament also split. As the higher, dynamic part of the filament reached a height of approximately 100~Mm, flare ribbons and two large EUV dimmings \citep{Thompson:2000} that spanned supergranular boundaries formed on the low solar atmosphere, indicating a successful eruption of this portion of the filament. In addition to the brightenings on the surface, brightenings that appear to trace the outside of the magnetic structure suspending the filament are observed during the eruption, as shown in the online movie associated with Figure~\ref{fig:splitting}. The remaining portion of the split filament is visibly perturbed at this point but is unable to successfully erupt, ultimately reforming a part of the original filament a few hours later. Therefore the part of the filament that we have focused on in this study, and used to define the eruption of the filament, is the dynamic portion that successfully erupts into the heliosphere at 05:53~UT on 11 December 2011.

\section{Observations}
	\label{S-observations}

SDO/AIA is an EUV imager that observes the Sun in 10 passbands, seven EUV channels (94, 131, 171, 193, 211, 304, 335~\AA), two FUV channels (1600, 1700~\AA), and one visible channel (4500~\AA) with a spatial resolution of 1.5''. The 171~\AA\ and 193~\AA\ passbands in particular have peak emission temperatures of 0.6 and 1.3~-~2~MK respectively and thus primarily observe the upper chromosphere and low corona. The HMI is another instrument on SDO designed to study the evolution of the photospheric magnetic field. HMI images the Fe~{\sc i} absorption line at 6173~\AA, with a spectral resolution of five points across the wings and core, and characterises the effect of Zeeman splitting on emission originating at $\approx$~100~km above the photosphere \citep[][]{Fleck:2011}. This produces an estimate of the LOS component of the photospheric field.

The STEREO mission consists of two identical non-Earth-orbit spacecraft. STEREO-A and STEREO-B are travelling ahead of and behind the Earth, respectively, and provide stereoscopic observations of the solar environment and heliosphere. Both STEREO spacecraft are equipped with an EUV imager that observes the Sun in four EUV wavelengths. The 195~\AA\ passband used here has a characteristic temperature of $\approx$~1.6~MK and images the upper chromosphere and hot plasma with a spatial resolution of 3.5''.


A movie detailing the evolution of the filament in EUV as seen by SDO and STEREO-B is included in the online version of this manuscript and associated with Figure~\ref{fig:splitting}. Due to an additional (unrelated) simultaneous eruption in the northern hemisphere but behind the limb, it was not possible to disentangle the two co-temporal CMEs in images from the \textit{Large Angle Spectrometric Coronagraph} \citep[LASCO;][]{Brueckner:1995} instrument on board the \textit{Solar and Heliospheric Observatory} \citep[SOHO;][]{Domingo:1995}.

\section{Results} 
  \label{S-results}

\subsection{Evolution of Magnetic Flux in the Filament Channel}
  \label{S-flux}

\begin{figure}
	\centerline{
    			\includegraphics[width=1\textwidth,clip=, trim= 43 0 10 0]{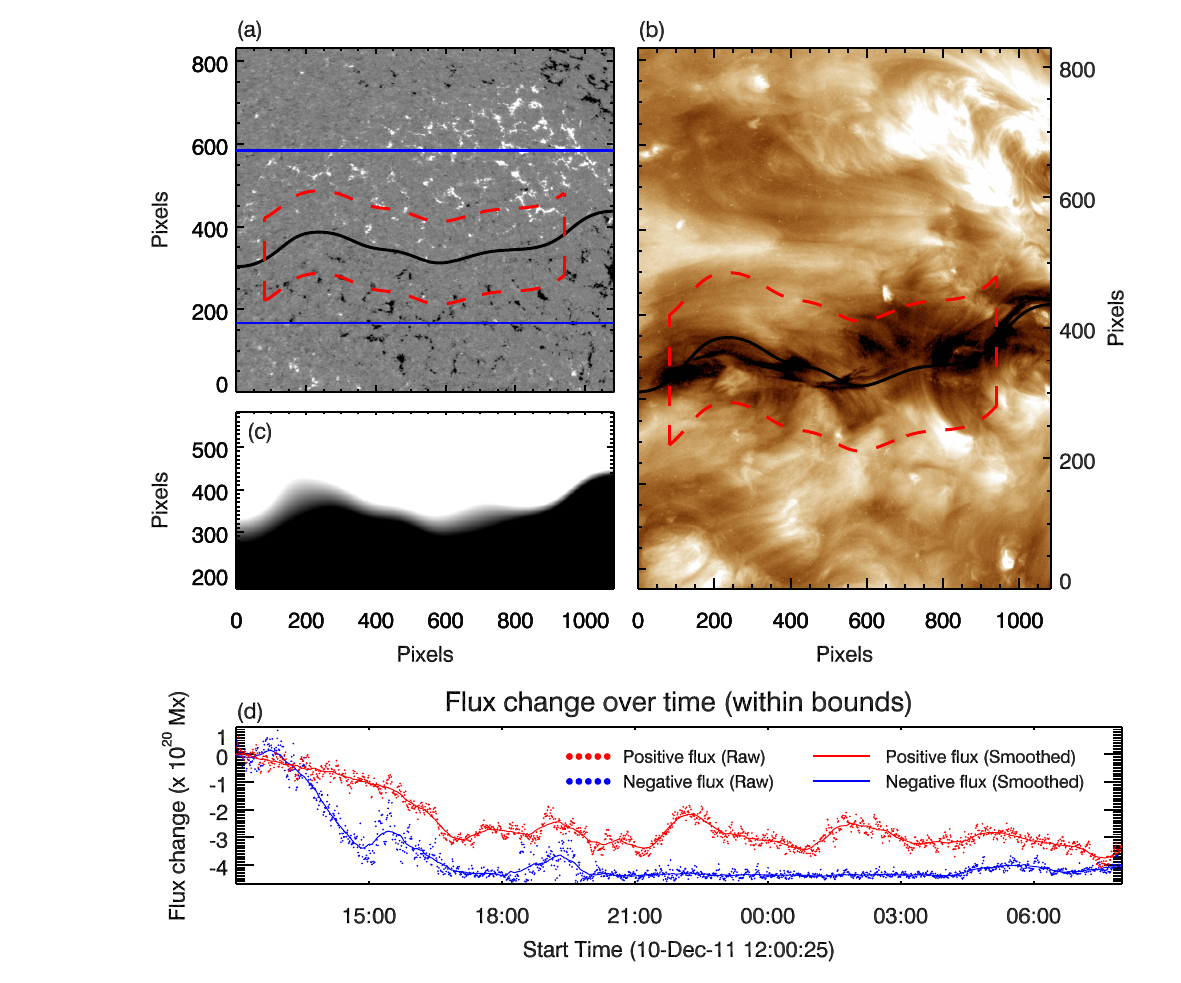}
               }
	\caption{The magnetic field evolution within the filament channel. \textit{Panel a}; HMI LOS magnetogram rotated to disc center. Solid blue lines show the bounds of the LOS magnetogram smoothed to define the position of the PIL as in \textit{(c)}, solid black line shows the position of polarity inversion line (PIL) based on smoothing regime, dashed red lines define the bounds of summation defined by features associated with the filament observed in AIA 193~\AA\ and 171~\AA\ passbands, as in \textit{(b)}. \textit{Panel b}; AIA 193~\AA\ passband image corresponding to same FOV as the HMI image in \textit{(a)}, used to define the region enclosed by the red dashed line. Image has been saturated to emphasise filament material. \textit{Panel c}; Result of the smoothed HMI LOS magnetogram that defines the location of the PIL. \textit{Panel d}; The evolution of flux contained within the boundaries defined in \textit{(b)}. Cancellation is present within the specified bounds until $\approx$~18:00~UT, after which the trend plateaus and remains near-constant.}\label{fig:pilsumlocation}
\end{figure}

Tracking the evolution of magnetic flux along the filament channel can give an insight into the evolution of the magnetic environment around the filament of interest. This can be done by isolating the area the filament channel occupies in HMI data. To do this, the HMI field-of-view (FOV) was selected to include the estimated location of the PIL of the filament and a small amount of the positive and negative region either side. The image was then successively smoothed by 1000 iterations of a 7.5 pixel width window to reduce the noise in the FOV. This approach was taken as restricting the saturation to $\pm$1~G was found to produce an image with a large noise component, making it impossible to get a location for the PIL. The iterative approach results in an image with a clear neutral line indicating a PIL, as shown in panel b of Figure~\ref{fig:pilsumlocation}, despite a very diffuse PIL in the LOS data. This was visually checked against the location of the filament in the SDO/AIA 193~\AA\ and 171~\AA\ passbands. The width of the region used to sum the magnetic flux was then set by identifying features connected to the filament channel using the 193~\AA\ and 171~\AA\ passbands. The upper bound was set to include the western footpoint and exclude the small cancelling bipole present at $\approx$~(650~px, 450~px) in Figure~\ref{fig:pilsumlocation}, and the lower bound was set at the latitude where the large mass deposit was seen to impact the surface. The east and west boundaries were defined by the appearance of the filament material. This ensures that as much as possible of the flux contained within the filament channel was included within the summation limits. However, this does not account for the inclusion of additional flux which might be unrelated to the filament. Similarly the boundary selection is a `by-eye' process, and assumes that the features chosen accurately represent the boundary of the material to be studied. A threshold of $\pm$30~G was applied to the FOV summation to reduce the noise in the result. This approach was then applied to all the corrected magnetic data throughout the observation period from 12:00~UT to 08:00~UT on 10 and 11 December 2011. This produces a value of summed positive and negative flux within the specified bounds for each time step. The results for the positive and negative flux variation were then smoothed over time using a 50 point average to subdue the small-scale variation. 

Panel~d of Figure~\ref{fig:pilsumlocation} shows a large decrease in flux (interpreted as cancellation) present within the specified bounds at the beginning of the observation period. This decrease corresponds to a value of $\approx$~3.5~$\times$~10$^{20}$~Mx of unsigned flux. The flux cancellation along the PIL is then seen to plateau after $\approx$~18:00~UT on 10 December 2011 and remains nearly constant for the rest of the observation period up to and after the eruption.%

\subsection{Bipole Emergence}
  \label{S-flux-emergence}

\begin{figure}
 \centerline{
 			\includegraphics[width=1\textwidth,clip=, trim=25 0 0 0]{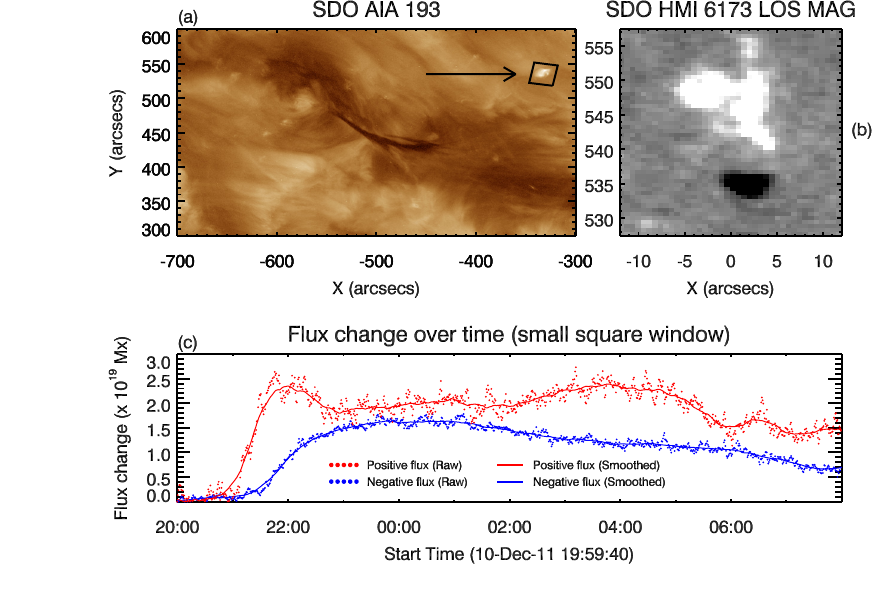}
            }
 \caption{The evolution of a small-scale emerging flux region at the edge of the filament channel. \textit{Panel a}; Location of the bipole emergence with respect to the filament of interest, as indicated by the arrow and black box \textit{Panel b}; The FOV, corresponding to the black box in \textit{(a)}, used for flux summation of the bipole emergence (positive=white, negative=black). Note that the bipole is surrounded by a positive-polarity magnetic environment. \textit{Panel c}; The evolution of the positive and negative flux contained within the small box surrounding the emerging bipole seen to the north-west of the filament. Emergence begins at approximately 21:00~UT on 10 December 2011 and negative flux, associated with the emergence only, peaks at 00:30~UT on 11 December 2011. Both AIA and HMI images have timestamps of 00:41~UT.}\label{fig:bpemerge}
\end{figure}

	Bright, low-lying loops observed in EUV close to the edge of the filament channel were seen to form several hours prior to the eruption on 11 December 2011 at 05:53~UT. The corresponding photospheric signature of this flux emergence was identified in LOS magnetograms as opposite polarity signatures growing and separating (forming a bipole) within the positive region of the diffuse bipolar region hosting the filament. The orientation of the bipole, observed using HMI, was such that its negative polarity was closest to the PIL of the region. Such adjacent, oppositely signed flux is a configuration established to be favourable for reconnection between the two systems \citep{Feynman:1995}.

	Figure~\ref{fig:bpemerge} shows the evolution of the flux contained within a small box surrounding the region into which the bipole emerged. The region contained within the FOV was located at -32$\degree$ longitude and +33$\degree$ latitude from Sun centre at 21:00~UT. The prepped data were de-radialised and de-rotated to allow the radial component of the magnetic field to be estimated, and to rotate the region-of-interest to disc center respectively. This placed the de-rotated region of interest at 0$\degree$ longitude and +33$\degree$ latitude. The FOV was then restricted to $\pm$12'' in x and 528--557'' in y, as seen in Figure~\ref{fig:bpemerge}b, and the sum total of the LOS magnetic fields with $|B|>30$~G within the enclosed area were calculated, with this process repeated for all time steps. The result was then smoothed by a moving 50-point average to subdue the small-scale variations and isolate just the overall trend. At 20:00~UT, in Figure~\ref{fig:bpemerge}, the values of the positive and negative flux were set to 0 to isolate the emergence of the bipole. 

	The emergence began at approximately 21:00~UT on 10 December 2011, indicated by the increase in both positive and negative flux in Figure~\ref{fig:bpemerge}. It is noticeable that the negative flux increased at a slower rate than the positive flux. Due to the predominance of positive polarity in the region of emergence, this lag may be due to the emerging negative flux interacting almost immediately with its surroundings and either remaining below the $\pm$30~G threshold or cancelling entirely. Alternatively, it could be an artifact of the simple assumption used in the calculation of the radial component, therefore introducing a flux imbalance in the photospheric field. It is also clear that the evolution of the positive flux and negative flux were non-identical. The FOV into which the bipole emerges was sufficiently isolated with respect to nearby flux that no flux breached the boundary of the FOV during the observation period. Despite this, the positive component displays sporadic variations suggesting that pre-existing polarity within the $\pm$30~G threshold was breaching and receding through this threshold over time. This indicates that the trend shown by the negative flux is more representative of the emergence of just the bipole into the FOV as its flux evolution only corresponds to that of the emergence. The negative flux can be seen to peak at approximately 00:30~UT on 11 December 2011 with a value of $\approx$~1.6~$\times~10^{19}$~Mx of unsigned flux emergence recorded.

\subsection{Morphological Analysis of Flows} 
  \label{S-flows}

\begin{figure}
 \centerline{
 			\includegraphics[width=1\textwidth,clip=]{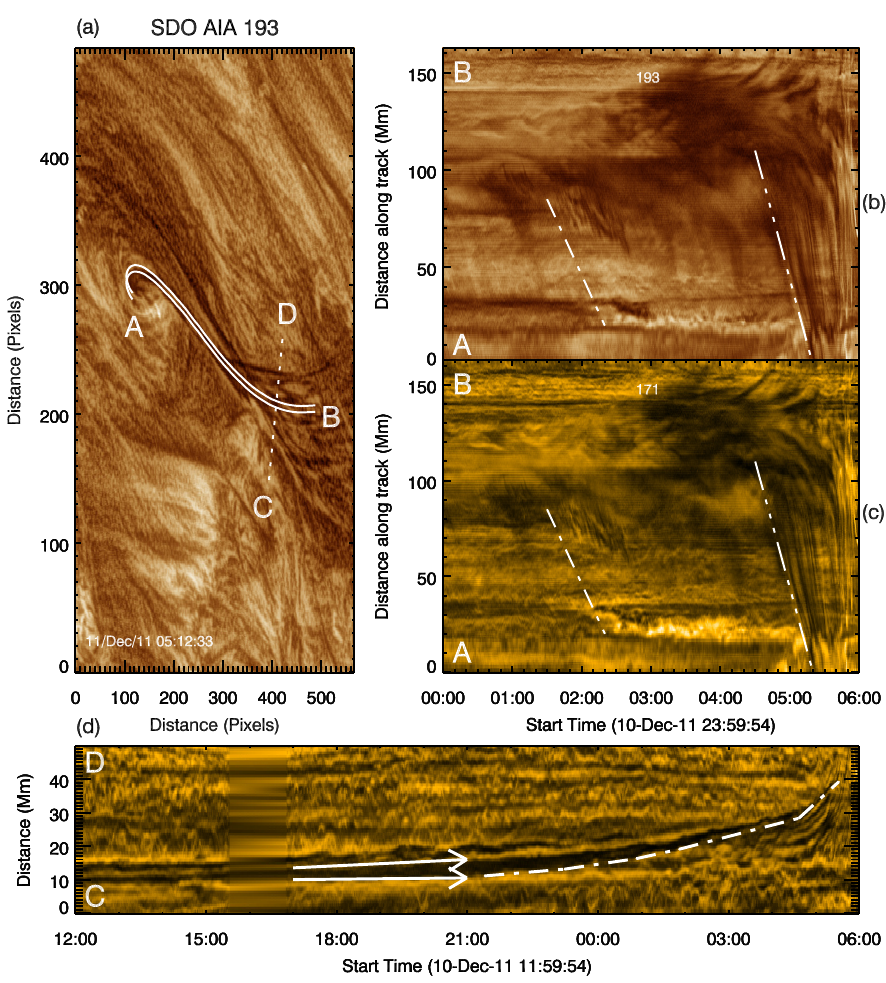}
            }
 \caption{Tracking intensity variations along and perpendicular to the filament axis. \textit{Panel a}; The de-rotated FOV used to specify the vectors that trace the motions of interest. The region contained within the white solid lines (A-B) was used to construct the stack-plots shown in panels \textit{(b)} and \textit{(c)}. The dotted white line (C-D) indicates the location of the line used to construct the stack-plot shown in panel \textit{(d)}. \textit{Panel b}; The temporal variation in pixel intensity, averaged across width of separation of the two lines, along the axis of the flows over time for the 193~\AA\ passband. \textit{Panel c}; The same as \textit{(b)} but for the 171~\AA\ passband. Dash-dot and dash-dot-dot-dot lines indicate start of the initial and the large mass-unloading. \textit{Panel d}; The temporal variation of pixel intensity along the dotted white line (C-D) in \textit{(a)}. Passbands were processed using the MGN technique to isolate the fine structure of the flows.}\label{fig:easttrack}%
\end{figure}

	Flows are seen by SDO/AIA to propagate away from the apex of the dynamic portion of the filament just prior to the eruption. As the flows propagated away from the apex, they appear to travel along a previously unidentified path that led away from the filament. We define this new path as the axis of the flows. Although the flows are seen as dark `blobs', images taken using SDO/AIA are sensitive to temperature variations in the emissive source. Thus any intensity variations observed could be temperature variations of the observed plasma environment or physical density variations and therefore plane-of-sky motions of the plasma itself. In order to distinguish the nature of the intensity variations as flows, the motions need to be temperature independent. By plotting the intensity of pixels along a vector against time it is possible to isolate how these intensity variations are evolving along the given vector; we refer to these as stack-plots. The stack-plots shown in panels b and c of Figure~\ref{fig:easttrack} detail the evolution of intensity variations recorded along the static solid, white vector specified in panel a of Figure~\ref{fig:easttrack}. The location of the vector was specified by hand-clicking along the path that the largest intensity variation is seen to have taken from the apex of the filament to the eastern footpoint just prior to the partial eruption of the filament. From the initial hand-clicked vector, a width was specified of two pixels either side to allow an average to be taken across the separation. This average was introduced to reduce noise in the recorded value of pixel intensity and to increase the signal of the potential flows against the background of the solar surface. Panels b and c of Figure~\ref{fig:easttrack} show the temporal variation in intensity along the path to the apex from the eastern footpoint of the filament for the 193~\AA\ and 171~\AA\ passbands, respectively. %
 
   The first large intensity variation observed to travel from the apex of the filament to its eastern footpoint occurred at $\approx$~01:30~UT. This is seen in panels b and c of Figure~\ref{fig:easttrack} as a darker feature originating at $\approx$~85~Mm along the track and impacting the surface at 02:30~UT, as indicated by the brightenings at $\approx$~30~Mm. As the vector was specified along the path taken by the largest intensity variation just prior to eruption, and given that filaments are highly dynamic structures, it does not fully trace the path of all intensity variations. It is however appropriately placed to record the initial movement from the apex, and the final movement and impact of the intensity variations at the surface. The surface brightenings are present throughout the lead-up to the eruption, between 20 and 30~Mm from 02:30 to $\approx$~05:00~UT in Figure~\ref{fig:easttrack}, suggesting the process causing the intensity variations decreased in extent but did not cease. The intensity variations were then seen to darken and dramatically expand from the filament centre to the surface, approximately one hour before eruption, noted in panels b and c of Figure~\ref{fig:easttrack} by the grouping of linear streaks angled towards 0~Mm \textit{i.e.,} the eastern footpoint. This larger motion then continued throughout the final hour leading up to the eruption. Intensity variations were also observed to propagate from the apex of the filament to its western footpoint, however these were far less intense or dynamic. These intensity variations on the western side remained constant after their initiation at 02:50~UT, persisting until the eruption of the filament, and simply served to highlight the location and large size of the western footpoint.
   
   In addition to the stack-plots made from the vector A-B in Figure~\ref{fig:easttrack}, the second vector (C-D) is a perpendicular bisect of the length of the filament. The change in the orientation of the dynamic portion of the filament over time is presented in panel d. Between $\approx$~17:00 and 21:00~UT on 10 December the filament can be seen to widen, as highlighted by the arrows. After the expansion of the filament, the entire filament appears to undergo a bulk, anti-clockwise rotation that persists up to the partial eruption of the filament at $\approx$~05:53~UT on 11 December.
   
   Figure~\ref{fig:easttrack} shows that the observed flows were spatially co-located in both the 171~\AA\ (0.63~MK) and 193~\AA\ (1.3~-~2~MK) passbands. This indicates that the variations were not temperature sensitive but were physical density variations showing the motions of material within the filament. As these intensity variations were indeed material motions, we can now consider the density of these flows with respect to the density of the rest of the filament structure.
 
\newpage
\subsection{Density Evolution}
	\label{S-density}

    Cool, dense chromospheric material \corr{that is} suspended in the hot, tenuous corona (i.e. filament material) appears in absorption in extreme ultraviolet (EUV) wavelengths \corr{below the Lyman continuum limit at 912~\AA}; photons are removed from the LOS predominantly by photoionisation \citep{Williams:2013}\corr{, so the efficiency of this removal is a function of wavelength}. \corr{The temperature of this material, however,} is low enough to assume there is negligible emission occurring at these wavelengths \citep[][]{Landi:2013}. In this case, the optical depth of the material ($\tau$) is defined by the column number density $N$ multiplied by the cross-sectional area of photoionisation $\sigma$,
\begin{equation}\label{opdep}
\tau~=~N~\sigma(\lambda) , 
\end{equation}
which will reduce the intensity of radiation passing through the material as,
\begin{equation}\label{iob}
I_{obs}~=~I_b\exp{(-\tau)} ,
\end{equation}
where $I_{obs}$ is the final observed intensity and $I_b$ is the intensity before passing through the material (`background').
The cross-sectional area of hydrogen and both neutral and singly-ionised helium is very similar at wavelengths below 227~\AA ~when weighted by the solar chemical abundances given by \citet{Grevesse:2007} (A$_H$~=~1, A$_{He}$~=~0.085), allowing the column number density of hydrogen to be calculated from the total optical depth,
\begin{equation}\label{hcolden}
N_H~\geq~\frac{\tau_{tot}}{2A_{He}\sigma_{HeII}}
\end{equation}
\citep[see][for a rigorous derivation.]{Williams:2013} 

The total optical depth of such material may be estimated provided the `background', or rather the unattenuated radiation field\footnote{`unattenuated radiation field' refers to the emission from behind and in front of the material in question; the radiation field as it would appear to an observer in the absence of the material would not be reduced in intensity by any absorption as there is no material there to absorb, or `attenuate' the background intensity.}, can be reasonably approximated. This may be done for highly dynamic material by taking an image co-spatial to the examined material some moments in time before or after the material is in that particular FOV. For less dynamic material, the background could be estimated from surrounding areas which are unobscured by cool, dense material. Therefore, provided two suitable images exist (one of the material to be measured, and one to estimate the unattenuated field), a lower limit on the hydrogen column number density may be calculated.

\subsubsection{Polychromatic Method}
	\label{S-Poly}

\begin{figure}
	 \centerline{
               \includegraphics[width=1\textwidth,clip=, trim=220 40 40 280]{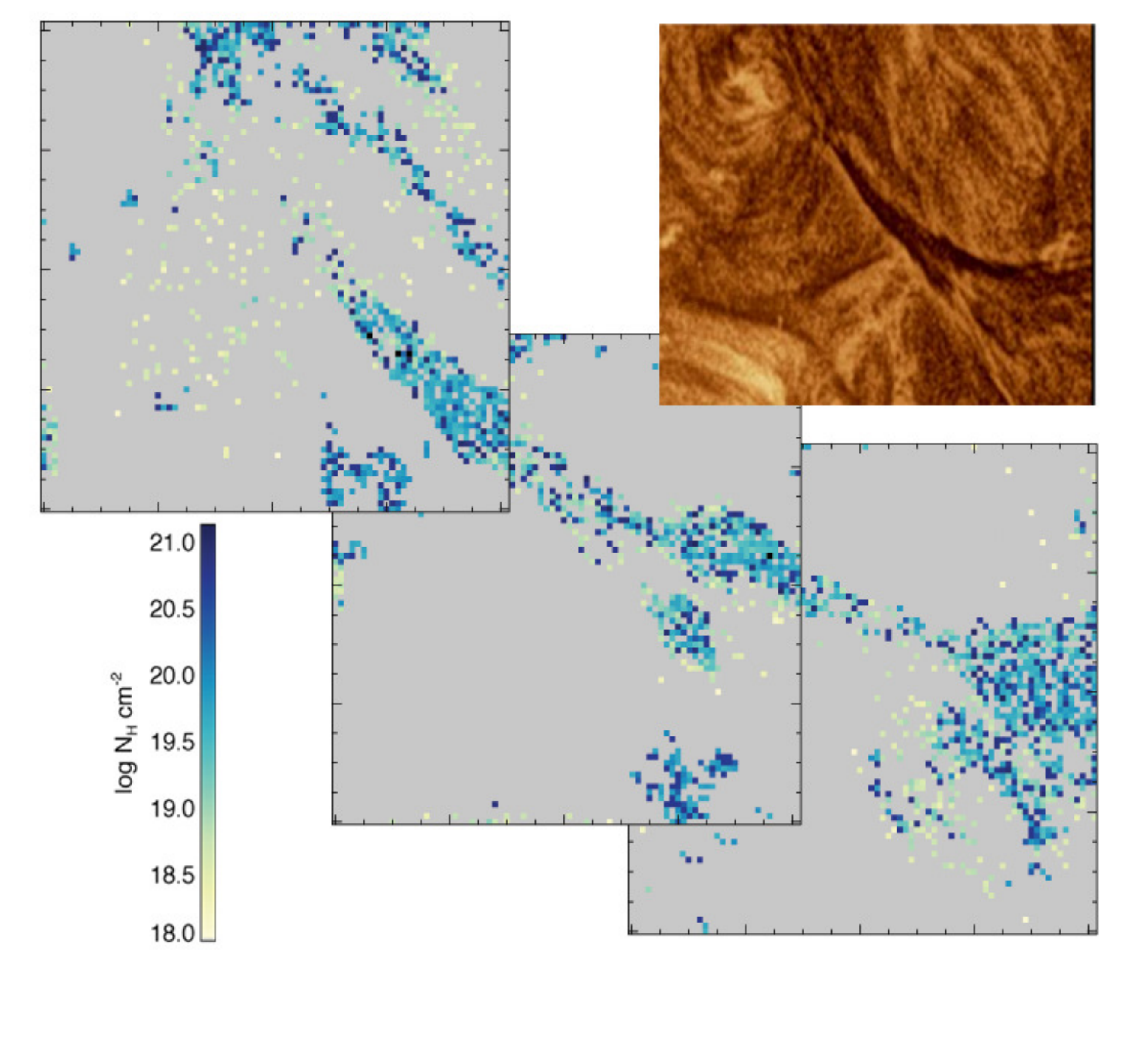}
               }\vspace{-0.04\textwidth}
     \caption{The mosaic of column density measurements of the dynamic portion of the filament from SDO/AIA captured at 01:51~UT. The mosaics have a 48''~$\times$~48'' FOV, with each tick separated by 3''. \textit{Top right}; Intensity image from SDO/AIA 193~\AA\ with the filament as seen at 01:51~UT.
     }\label{fig:sdo_density}
\end{figure}

\begin{figure}
	 \centerline{
     			\includegraphics[width=1\textwidth,clip=, trim=0 -20 0 100]{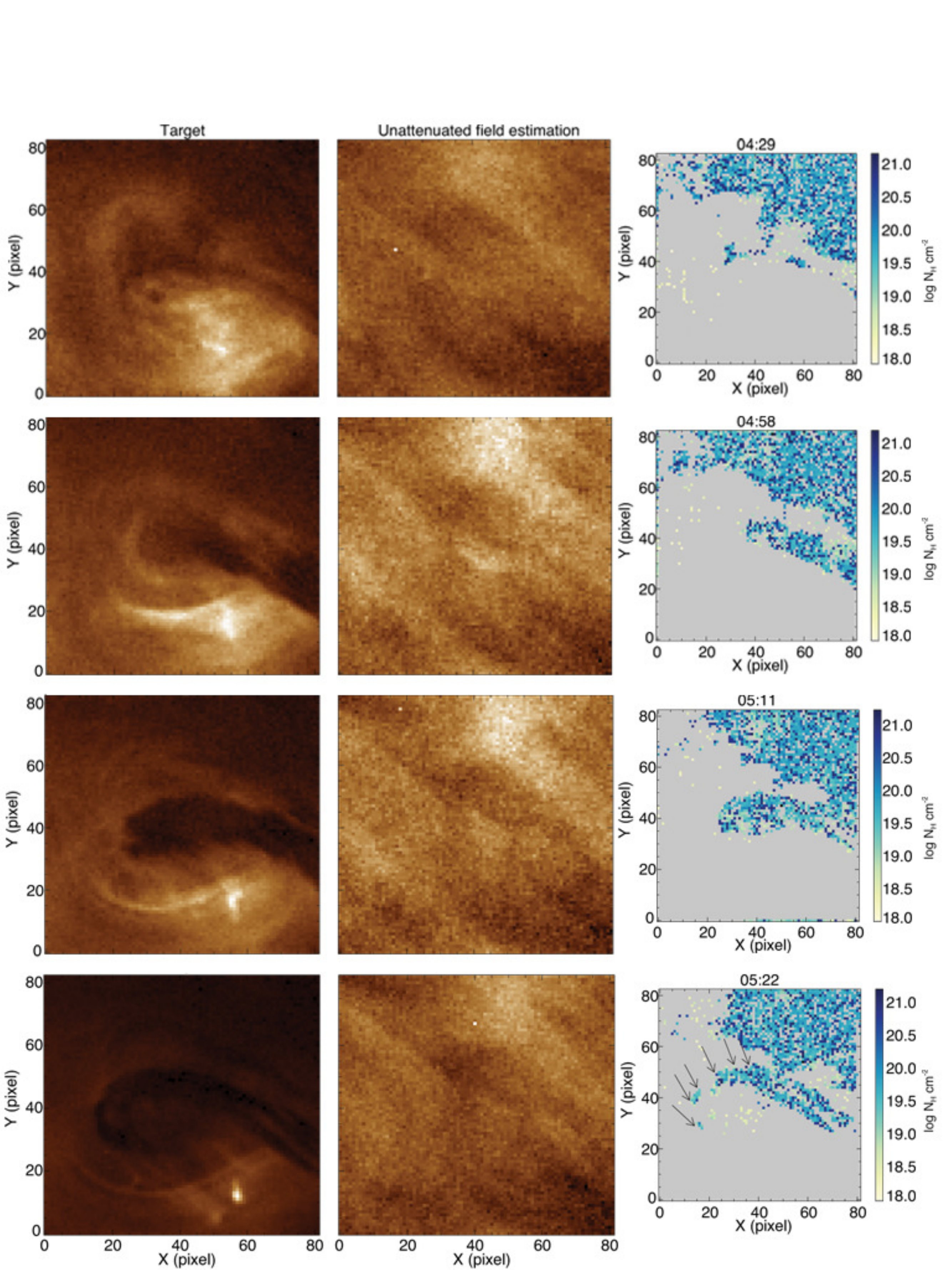}
              }
     \caption{Density evolution of the large mass-unloading between 04:29 and 05:22~UT on 11 December 2011, using the polychromatic method applied to SDO/AIA data. \textit{Left}; FOVs used in the density determination method that include the footpoint of the large mass-unloading, defined in \textit{(a)} of Figure~\ref{fig:context}. \textit{Middle}; Unattenuated radiation field estimation for the intensity image shown in the \textit{left}. \textit{Right}; Result of the density determination method based on the specified target and unattenuated field estimate. The hooked shape of the dense object at 05:22~UT, as indicated by the arrows, is due to the material approaching the surface via curved field lines. All times indicated above the panels are in UT.}\label{fig:sdo_density_motion}
\end{figure}

	As previously mentioned, should the filament material be observed in $\geq$~3 wavelengths below 227~\AA\ (the cross-section of ionisation limit for He~{\sc ii}), the optical depth can be used to constrain a model which includes the fraction of emission; the unattenuated radiation field includes not only background radiation but also emission from hot coronal material between the filament material and observer. Furthermore, fine structuring in the filament material may allow background emission to pass through unobstructed, and as such a pixel-filling factor should be considered. Therefore, the intensity observed is given by,
\begin{equation}\label{iobs}
I_{obs}~=~I_b(f\exp{(-\tau)}~+~(1~-~f))~+~I_f ,
\end{equation}
where $f$ is the pixel-filling factor (\emph{i.e.} the fraction of each pixel occupied by material) and $I_f$ is the foreground emission. Rearranging, we have.
\begin{equation}\label{thed0}
1~-~\frac{I_{obs}}{I_b~+~I_f}~=~f\frac{I_b}{I_b~+~I_f}~(1~-~\exp{(-\tau)}) ,
\end{equation}
where the unattenuated radiation field is approximately equal to \corr{$I_b~+~I_f$} (emissivity blocking, the emission which would be emanating from the hot corona in the location of the filamentary material were it absent, is negligible due to the small volume relative to the rest of the corona), and so the left-hand-side of Equation~\ref{thed0} is measurable, denoted as $d(\lambda)$. On the right-hand-side, a substitution can be made,
\begin{equation}
f\frac{I_b}{I_b~+~I_f}~=~G,
\end{equation}
This reduces the model to have two free parameters, and so if $d$ is measured in three or more wavelengths, the model may be constrained by \emph{e.g.} a least-squares fit. Although the multiple wavelengths below 227~\AA\ required by this technique are captured by SDO/AIA, from this point-of-view (POV) the unattenuated radiation field is not only more dynamic but also more structured. This introduces uncertainties in the estimated radiation field and hence the calculated density. For more detail on this method, see \citet[][]{Heinzel:2008}, \citet[][]{Williams:2013}, and \citet{Carlyle:2014a}.

	 In Figure~\ref{fig:sdo_density} we study the density of the dynamic portion of the filament. A mosaic is presented as computing the column density for a square of this size would be much more computationally demanding, and given that the corners are of no interest, this was deemed unnecessary. Furthermore, unattenuated estimation assumption is more difficult to satisfy as the Sun is highly dynamic and structured, so using three similar, smaller frames increases the reliability of the results. The requirement for an unattenuated background field was well satisfied at 01:51~UT, which is why this time was chosen for the analysis. 
     
      In Figure~\ref{fig:sdo_density_motion} we focus on the motion of the large density flow from the filament apex to the eastern footpoint. As before, a sufficiently small FOV was chosen and the result of applying the density determination method to this FOV at multiple times is presented, revealing the evolution of density in the frame throughout the mass-unloading. Material is clearly seen to enter the FOV from the apex of the filament in the centre-right of each image, and follow a curved path corresponding to the curved field lines from the filament apex to the surface.

	The polychromatic technique applied to both the entire dynamic portion of the filament, and the large mass-unloading as seen in the SDO/AIA data returns a mean column number density of roughly 1~$\times$~10$^{20}$~cm$^{-2}$.

\subsubsection{Monochromatic Method}  
	\label{S-Mono}
    
\begin{figure}
	 \centerline{\hspace*{-0.025\textwidth}
               \includegraphics[width=0.36\textwidth,clip=, trim=0 73 100 20]{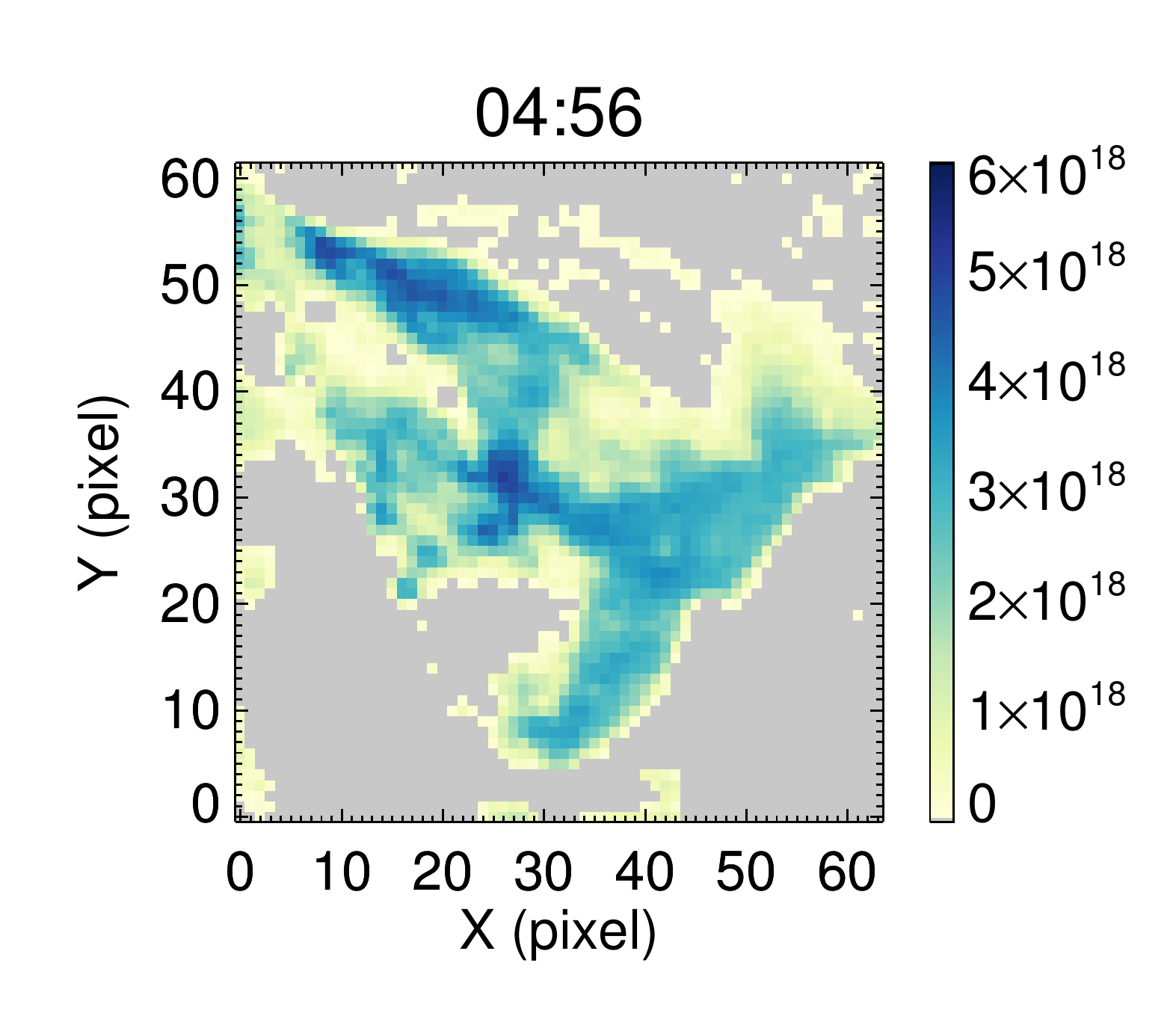}
               \hspace*{-0.025\textwidth}
               \includegraphics[width=0.27\textwidth,clip=, trim=88 73 100 20]{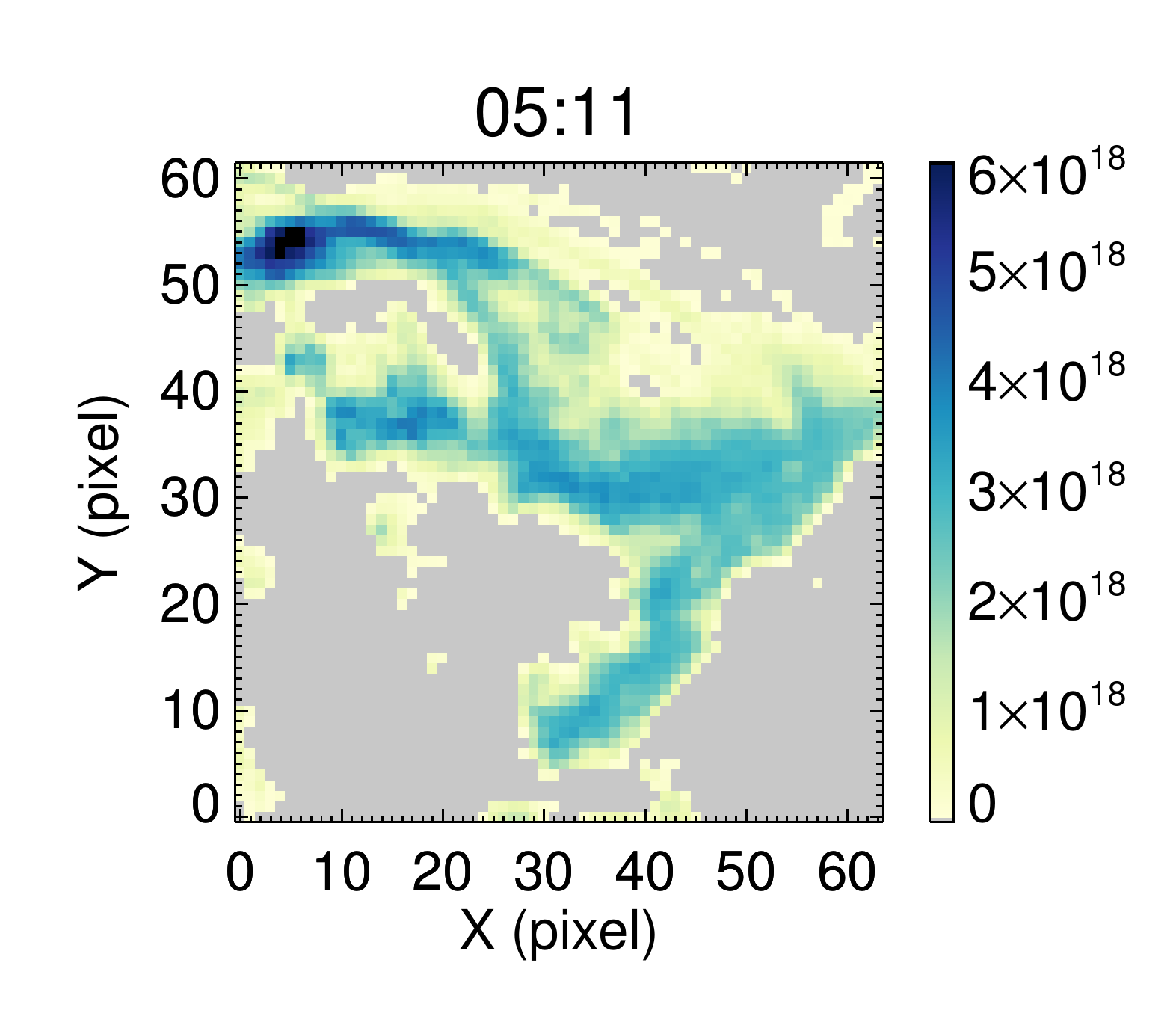}
               \hspace*{-0.025\textwidth}
               \includegraphics[width=0.37\textwidth,clip=, trim=88 73 0 20]{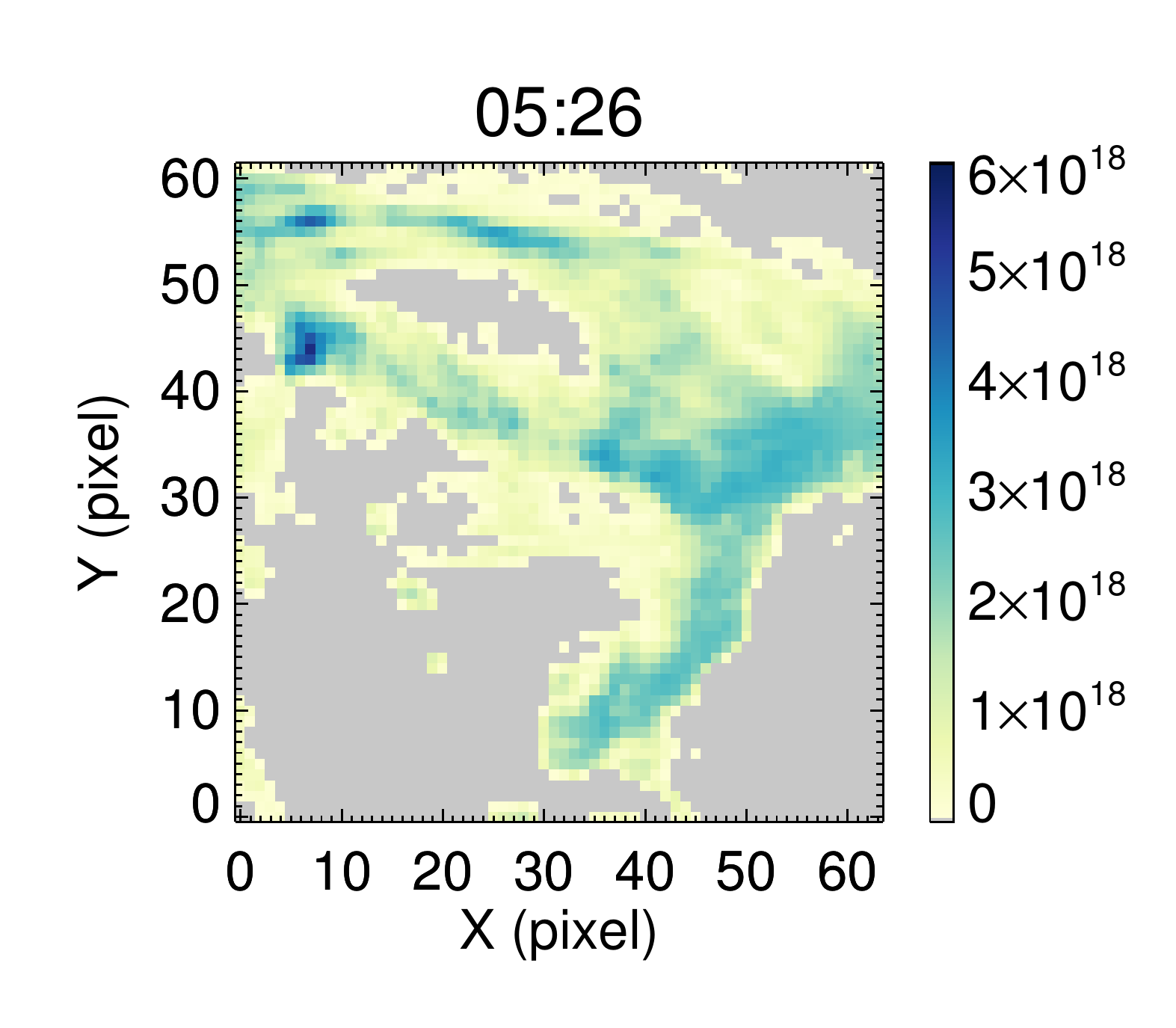}
              }\vspace{-0.06\textwidth}
     \centerline{\hspace*{-0.025\textwidth}
               \includegraphics[width=0.36\textwidth,clip=, trim=0 20 100 0]{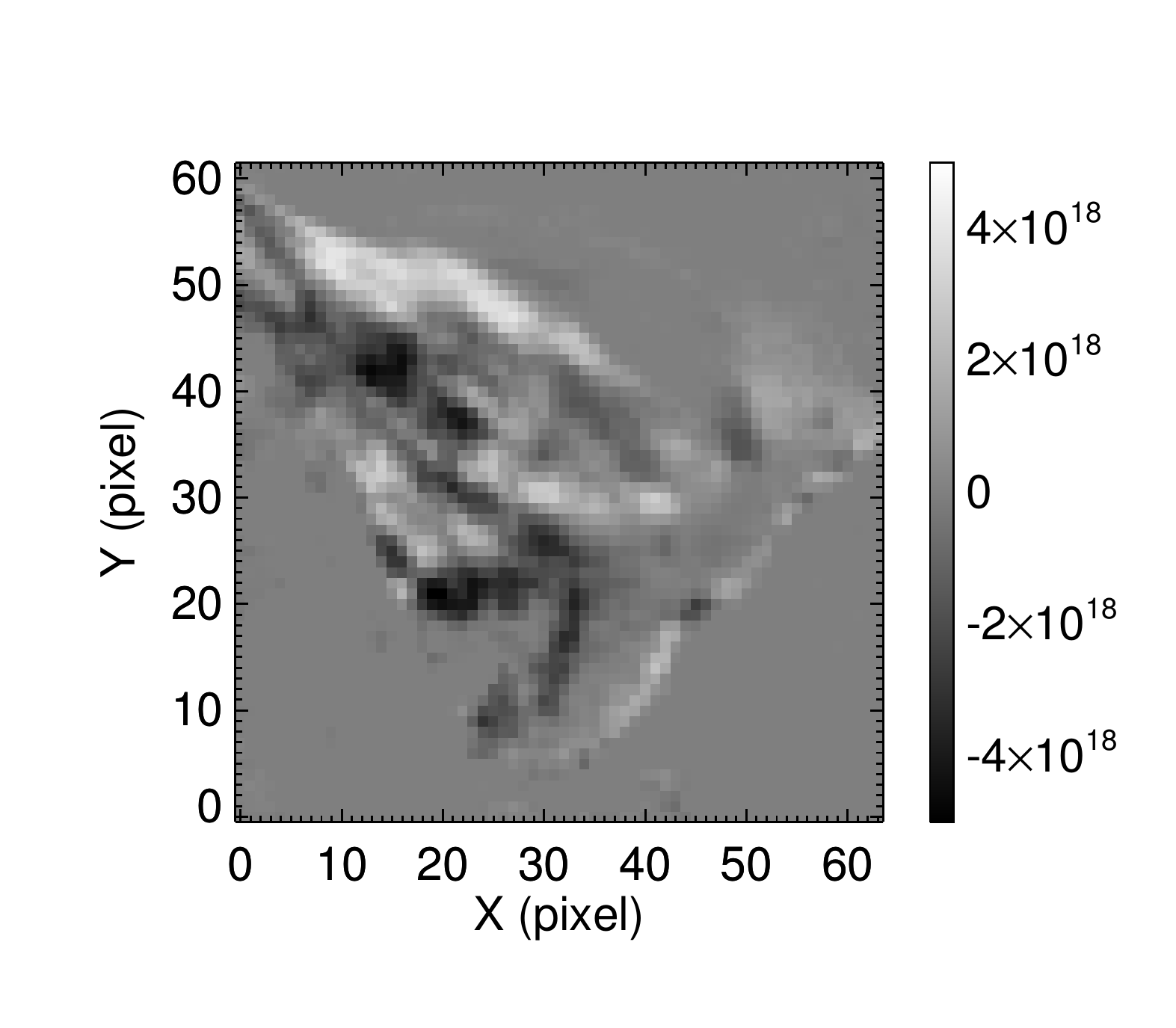}
               \hspace*{-0.025\textwidth}
               \includegraphics[width=0.27\textwidth,clip=, trim=88 20 100 0]{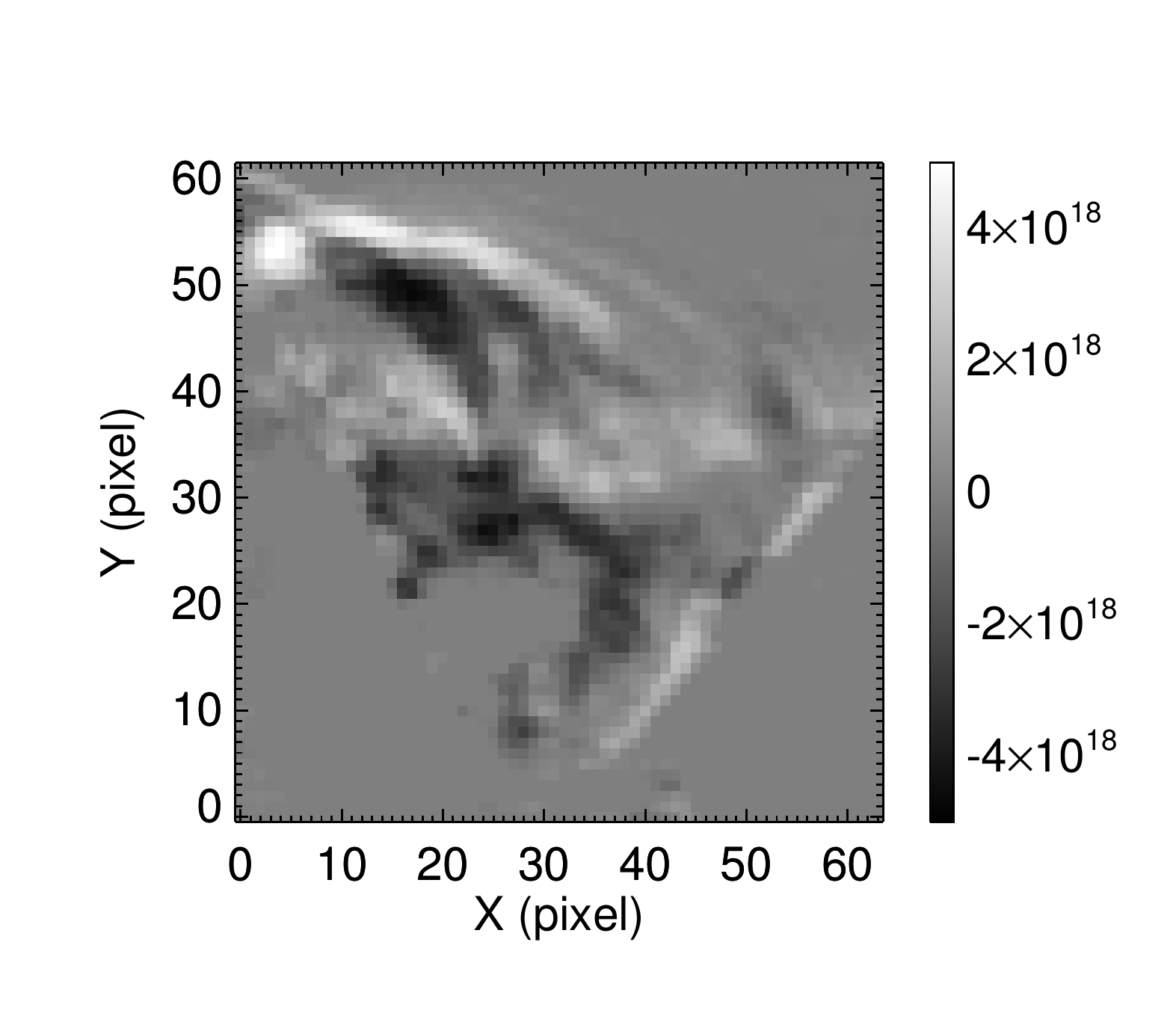}
               \hspace*{-0.025\textwidth}
               \includegraphics[width=0.37\textwidth,clip=, trim=88 20 0 0]{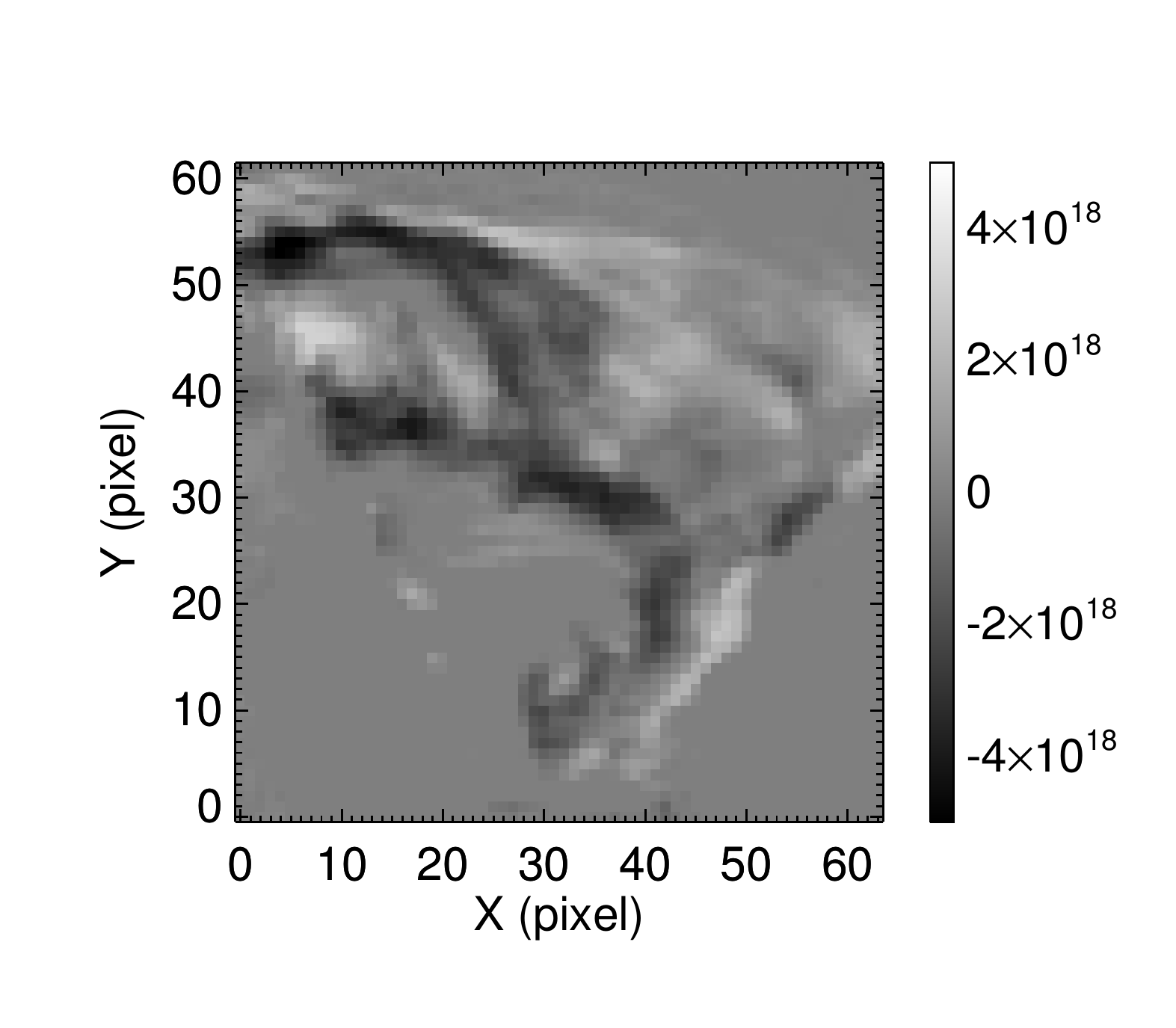}
              } 
     \caption{The evolution of column number density and flows in the dynamic portion of the filament as seen from STEREO-B/EUVI. \textit{Top row}; Evolution of column number density, measured in N$_H$ cm$^{-2}$, detailing the mass-unloading initiated at $\approx$~04:40~UT on 11 December 2011. The large mass deposit can be seen as an increase in column number density at the top-left of the image. \textit{Bottom row}; Running difference images of the column number density evolution measured in $\Delta$N$_H$ cm$^{-2}$. Features getting more (less) dense in time appear white (black). This further highlights that the mass is moving towards the north-east footpoint of the filament (top-left of each image).}\label{fig:stereo_mass_density}
\end{figure} 

    On 11 December 2011, STEREO-B was ideally positioned to view the erupting filament at a near perpendicular LOS with an angle $\theta_{STEREO-B}=108\degree$ to the Sun--Earth line. This provided a rare opportunity to view an eruption contemporaneously both on-disc (SDO/AIA) and off-limb (STEREO-B/EUVI), and help disentangle the structure of the filament that would be otherwise unachievable with single-perspective observations. From the POV of STEREO-B, the filament was projected against a slow-changing background of the corona, and as such the unattenuated background radiation field was well approximated by using an exposure taken in the location of the filament following its eruption. This method was only applied to data collected by the 195~\AA\ passband on board STEREO-B/EUVI only as the cadence for the 171~\AA\ channel was too low to be useful here. Due to having access to a single wavelength for these observations from STEREO-B and therefore no estimation of a filling factor or fraction of emission, density determination is only possible using the simpler model of Equation~\ref{hcolden}, a monochromatic estimation. Unfortunately, not being able to include the filling factor or fraction of emission in the estimation results in a lower estimation of the column number density. The results of this analysis are summarised in Figure~\ref{fig:stereo_mass_density}.

	The monochromatic technique applied to the STEREO-B/EUVI data returns an average lower column number density limit of approximately 4.5~$\times$~10$^{18}$~cm$^{-2}$ for the filament, as seen in the top panels of Figure~\ref{fig:stereo_mass_density}. As before, applying this method to many snapshots over the course of the evolution of the filament highlights the evolution in column number density over time. These results show a gradual increase in total mass in the target frame from 4~$\times$~10$^{13}$~g to 8.8~$\times$~10$^{13}$~g over approximately 8 hours, consistent with the filament slowly rising into the FOV; the evolution of mass is shown in panel~c of Figure~\ref{fig:kinematics}. This suggests that the filament rising into the FOV and associated increase in measured mass has partially masked the initiation of the smaller-scale mass-unloading at 01:30~UT. This increase is followed by overdensities moving in the direction of the NE footpoints (highlighted by subtracting each column number density map from the next, creating a `running density difference' image as shown in the bottom panels of Figure~\ref{fig:stereo_mass_density}), before these overdensities appear to suddenly drain down towards the eastern footpoint, reducing the total target mass to 2.2~$\times$~$10^{13}$~g in just over an hour, as seen in panel c of Figure~\ref{fig:kinematics}.

	These results indicate that the filament studied here was large and dense throughout the lead-up to its eruption. A large and sudden decrease in the density of the material contained within the filament is apparent from 04:00~UT onwards to eruption, decreasing the total mass in its measurable portion by more than 2/3 ($\sim$~6.6~$\times$~10$^{13}$~g). 
\subsection{Filament and Plasma Kinematics} 
  \label{S-kinematics}

\begin{figure}
	 \centerline{
     			\includegraphics[width=1\textwidth, clip=, trim=0 45 0 10]{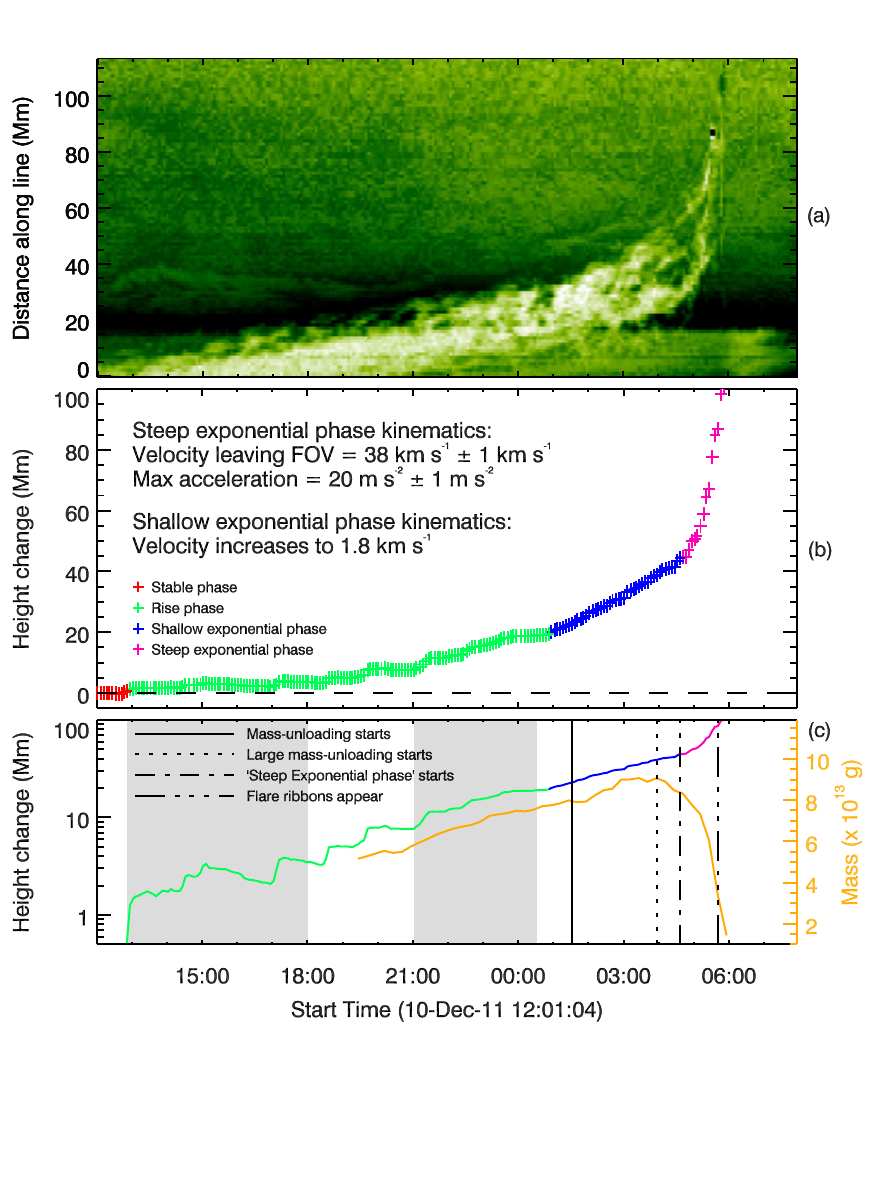}
     			} 
 \caption{The evolution of the filament height and mass with time. \textit{Panel a}; The height--time stack-plot taken along the line specified in Figure~\ref{fig:context}c using a reversed colour table. \textit{Panel b}; The solar-rotation-removed evolution of the dynamic portion of the filament as it evolved and erupted. Red indicates the stable phase, green indicates the rise phase, blue indicates the shallow exponential phase and magenta indicates the steep exponential expansion phase. Kinematics are derived from the fitting of an exponential function to the magenta region of evolution, and a line to the blue region. \textit{Panel c}; The h--t profile as in \textit{(b)} but with a logarithmic scaling, compared to the evolution of mass contained within the FOV enclosing the filament over time, as seen by STEREO-B/EUVI. The initial increase in mass corresponds to larger portions of the filament being visible above the limb over time due to solar rotation, and can therefore be considered to be artificial. Shaded patches indicate the times of flux cancellation recorded along the PIL and the nearby bipole emergence. Vertical lines indicate (from left to right) time of first visible mass-unloading, the time of the largest mass deposit, start of the steep exponential expansion, and initial appearance of the flare ribbons and twin EUV dimmings.}\label{fig:kinematics}%
\end{figure}

	In the previous sections we have identified several large, dense pockets of material moving away from the apex of the filament structure and down to the surface. The 1D velocity estimate, derived from the stack-plots in Figure~\ref{fig:easttrack}, of the large mass-deposit leaving the apex of the filament is $\approx$~28~km~s$^{-1}$. The second perspective offered by STEREO-B/EUVI shown in Figure~\ref{fig:stereo_mass_density} indicates that the large mass deposit originated from a height of $\approx$~40~Mm and $\approx$~30 minutes earlier than suggested in the stack-plots of Figure~\ref{fig:easttrack}. The combination of the stack-plot observations made from both spacecraft therefore suggest that the mass-deposit had a 2D linear acceleration of $\approx$~$12$~m~s$^{-2}$, reaching a velocity of $\approx$~31~km~s$^{-1}$ as the material approached the surface. The acceleration of the unloading mass is an order of magnitude lower than free-fall at the solar surface.
 
    As previously stated in Section~\ref{S-roadmap}, the dynamic portion of the filament erupts whilst the restrained portion does not. Panel~a of Figure~\ref{fig:kinematics} shows the height--time (h--t) evolution of the dynamic portion of the filament from the perspective of STEREO-B. The evolution of the filament described in the stack-plot is directly in the plane-of-sky, the eruption occurred at $\approx$~$90\degree$ to the Sun--STEREO-B line and is therefore assumed to be approximately radial. The evolution of the filament height with time was measured by selecting the leading edge of the filament by hand. This was repeated six times to minimise user bias and provide an average position. The contribution to the h--t evolution due to solar rotation was then removed, revealing four main evolution domains within the lead-up to the eruption and summarised in panel~b of Figure~\ref{fig:kinematics}: 
 
	1. A stable phase (red) in which it appears the only contribution to the stack-plot evolution was due to solar rotation. The average position of this phase is set as a distance of zero so that any subsequent h--t evolution refers to a deviation from the height of the filament during this stable phase.%
 
 	2. A rise phase (green) that describes the deviation from stable phase and includes the initial lift-off from the surface.
 
	3. A shallow exponential phase (blue), defined as the h--t evolution is approximately linear when plotted on a logarithmic scale as in panel c of Figure~\ref{fig:kinematics}.
    
    4. A steep exponential phase (magenta). Fitting the exponential phase of the h--t profile with the exponential function suggests the final radial velocity of the filament leaving the FOV was $v$~$\approx$~$38$ $\pm$~$1$~km~s$^{-1}$, a larger-than-average velocity for a quiescent filament eruption according to \citet{Loboda:2015}. An acceleration of $a$~$\approx$~$20$ $\pm$~$1$~m~s$^{-2}$ was found to follow the reduction in mass within the filament and is indicative of the initiation of the impulsive acceleration phase of a CME \citep[][]{Schrijver:2008}.
    
    Panel~c of Figure~\ref{fig:kinematics} compares the h--t profile of the filament using a logarithmic scale with the evolution of mass from the column number density measurements of Section~\ref{S-density}. It is shown that the expansion of the filament between $\sim$~01:00~UT and 04:40~UT is exponential, as indicated by the linear evolution. The unloading of mass is seen to begin at $\sim$~04:00~UT; after a large proportion of the mass had drained from the system the h--t profile is seen to have accelerated to a larger exponential expansion.%

\section{Discussion} 
      \label{S-discussion}      

    Filament (prominence) eruptions, a progenitor of CMEs, have been studied for many decades \citep[][]{Parenti:2014}. The myriad studies describing the magnetic destabilisation of a solar filament have greatly advanced our understanding of the different solar eruption triggers and drivers \citep[][]{Aulanier:2010}. However, whilst the role played by the material contained within the flux rope has been acknowledged by several authors \citep[\textit{e.g.,}][]{Fan:2017}, it is largely neglected, with the general consensus being that the material is unimportant \citep[\textit{e.g.,}][]{Torok:2005,Fan:2007,Torok:2011}. Here we present an event strongly indicating that the internal dynamics of the mass cannot be ignored when considering an evolving and destabilising flux rope.
      
    At the start of the period chosen for analysis of this event (12:00~UT on 10 December) a decrease in surface flux is recorded along the PIL that lies underneath the filament of interest. This indicates flux cancellation and the increase in concentration of non-potentiality along the PIL. Indeed, if there was a pre-existing flux rope present along the PIL, the negative trend of total flux implies the addition of flux to the flux rope system. The easternmost border of the summation bound in Figure~\ref{fig:pilsumlocation} lies at $\approx$~60$\degree$. Therefore, due to the increased noise in LOS magnetic data past the 60$\degree$ limit \citep[][]{Hoeksema:2014}, we cannot study the evolution in flux further back in time and can only infer the connection to the flux rope formation process via flux cancellation described in \citet{vanballegooijen:1989}. The plateauing trend in the flux evolution suggests the reconnection driving the formation of the proposed flux rope had ceased by or just after 18:00~UT on 10 December. This non-potential magnetic system would find a new equilibrium; for a flux rope this can be achieved through an expansion and associated height increase. It is possible that the widening of the filament (panel d of Figure~\ref{fig:easttrack}) is the observational signature of the filament rising towards the observer. Unfortunately, due to the data gap in the AIA 171 passband between 15:32 and 16:50~UT on 10 December it is not possible to extrapolate this backwards in time.
   
   The emergence of the nearby bipole studied in Section~\ref{S-flux-emergence} began at $\sim$~21:00~UT, after the plateauing of flux cancellation recorded along the PIL below the filament of interest. Work by \textit{e.g.,} \citet{Feynman:1995} and \citet{Chen:2000} suggests that this emerging bipole was preferentially oriented for reconnection with the field overlying the filament of interest \textit{i.e.}, the negative polarity of the bipole was closer to the negative polarity of the hosting bipolar field. Indeed, if there was an interaction between the two systems then the overlying field above the filament would have been weakened as a result of the proposed reconnection, as in \citet{Williams:2005}. The filament would then have been able to expand to a higher height within the corona at a speed proportional to the reconnection rate between the bipole and the field overlying the filament. Based on the stack-plot presented in panel a of Figure~\ref{fig:kinematics}, panels b and c of the same figure show that the filament appeared to have been in equilibrium for approximately an hour prior to 21:00~UT. After 21:00~UT and the beginning of the emergence of the bipole, the same plots describe the filament as having resumed its expansion through the corona. The h--t evolution of the filament is then seen to have plateaued at 23:30~UT. At this time, the nearby bipole was approaching the peak of its emergence (see Figure~\ref{fig:bpemerge}, actual peak 00:30~UT on 11 December) and therefore the reconnection rate between the two systems would have reduced and ultimately stopped. It is worth noting that an additional, smaller bipole is seen to have emerged at $\sim$~20:30~UT beneath the western portion of the filament. Observationally it appears there was also some re-organisation of the field topology in the vicinity of this flux emergence, as indicated by the sporadic, bright extensions away from the location of the bipole. However, the orientation of this smaller bipole with respect to the surrounding field suggests that it was not favourably oriented for reconnection, as was the case with the larger bipole that has been studied in Section~\ref{S-flux-emergence}. Nevertheless, it is possible that this smaller bipole was involved in the evolution of the system, even if to an unmeasurable degree. Whilst the presented analyses suggest a tenuous connection between the studied emergence and the evolution of the filament, the post-eruption large twin EUV dimmings associated with the footpoints of the proposed flux rope \citep[cf.][]{James:2017} are seen to have migrated counter-clockwise and the western dimming is seen to have approached and eventually enveloped the location of the studied bipole, indicating a relationship between the bipole and the footpoints of the erupting CME. However, it is unlikely that such small bipoles were the sole influences on the evolution of the filament height after 21:00~UT on 10 December.
   
   EUV and LOS magnetogram observations of the region surrounding the filament of interest suggest that the proposed flux rope containing the filament would have been left-handed with a negative helicity, as inferred by its location in the northern hemisphere, its roots in a positive-leading and negative-trailing diffuse bipolar region (panel b of Figure~\ref{fig:context}), and the shear angle of the associated loops. \citet{Green:2007} suggest a flux rope in this configuration should rotate anti-clockwise about its PIL as it expands. Panel d of Figure~\ref{fig:easttrack} is a stack-plot defined by a vector that is a perpendicular bisector of the axis of the dynamic portion of the filament. In this plot, at $\sim$~21:00~UT on 10 December, the filament is seen to have started rotating counter-clockwise about its PIL. The combination of this observation and the previously discussed expansion of the filament beginning at approximately the same time \corr{is consistent} with those conclusions presented in the paper by \citet{Green:2007}. In addition to this, the counter-clockwise rotation can be seen to continue throughout the period leading up to the eruption at 05:53~UT on 11 December, whereas the emergence of the bipole -- the possible cause of the height increase from 21:00~UT on 10 December -- ceased by 00:30~UT on 11 December. This suggests that the observed persistence in rotation is in fact due to the proposed flux rope becoming kink unstable. Nevertheless, this does not necessarily override the role of the emerging bipole in the expansion of the filament, as the combination of the two mechanisms are likely to have influenced the temporal evolution of the height of the studied filament.
   
   Up to this point we have referred to the magnetic structure that possibly contains the filament as being of the flux rope topology. However, the event described here lacks certain features that are usually identified in observations when a flux rope is in fact present, for example, there is little evidence of a cavity in SDO/AIA 193~\AA\ images that would outline the shape of a possible flux rope when passing the eastern limb. The only explanation to this is that the angle that the filament makes with the LOS of either SDO or STEREO-B is not optimal for a cavity observation, as we are not observing the structure along its axis \citep[cf.][]{Gibson:2006, Forland:2013}. However, the combined observations of flux cancellation in the filament channel, the suggested onset of kinking, the post-eruption twin EUV dimmings, and the brightenings along the length of the filament during the eruption that appear to outline helical field, most visibly seen in the movie associated with Figure~\ref{fig:splitting}, are most consistent with the flux rope theory.
   
   Interestingly, the filament can be seen to have fanned out during this kinking. This is most evident on the western side of the filament as the field lines associated with the fanning are highlighted by material suspended along their length, as previously stated in Section~\ref{S-flows}. In the movie associated with Figure~\ref{fig:splitting} it is possible to see that the flows trace the curvilinear paths of the fanning field lines from $\sim$~01:30~UT until eruption. As we have assumed a flux rope configuration of the magnetic environment surrounding and suspending the filamentary material, by studying the magnetic topology of simulations by authors such as \citet{Roussev:2003}, \citet{Mei:2017}, and \citet{Guo:2017}, it is difficult to reconcile how the highly twisted field lines of these studied flux ropes could produce the observed linear-like motions of plasma; plasma beta (plasma pressure / magnetic pressure) is low in the corona and therefore charged material is line-tied. Interestingly, extrapolations of less-twisted flux ropes, such as those by \citet{Su:2011} and \citet{Jiang:2014}, are easier to compare to the observations as they contain very weakly twisted field lines that pass through the axes of the extrapolated flux ropes. However, as pointed out by \citet{Aulanier:1998}, the location of the dips that contain the cool, dense material of the filament are unlikely to reach entirely up to the axis of the flux rope, with the majority of the material lying in the lower portions. Nevertheless, the assumption that the filament material lies entirely within the dips of the magnetic flux rope is a first order approximation. With the inclusion of the thermodynamic instability, theorised to occur in the solar environment \citep[][]{Field:1965}, material may be able to be suspended in higher portions of a flux rope including the more weakly twisted field of the flux rope axis. In addition, work completed by \citet{Su:2011}, and more recently by \citet{Polito:2017}, were able to successfully reconstruct the diffuse footpoints of flux ropes, as their observations had initially indicated, a feature not currently achievable through simulations. Therefore the combination of magnetic dips, the thermal instability, and the ability for a diffuse flux rope footpoint offer explanations to the observations presented in this paper, \textit{i.e.,} that the material can be seen to travel along curvilinear field lines associated with the flux rope, and have a rooting that spans supergranular boundaries.%

    Continuing with the evolution of the filament, Figure~\ref{fig:kinematics} shows that the proposed flux rope containing the filament became marginally unstable at 01:00~UT on 11 December, as denoted by the linear height--time evolution on the logarithmic scale of panel~c. As a consequence of flux rope expansion, flux rope field lines that suspend filament material above the surface increase in their gradient with respect to the surface. At some time, this would cause the concave-up sections of the field lines to become more shallow and even disappear and therefore no longer capable of supporting the filament material against gravity \citep[][]{Fan:2017}. Therefore, filament material within such a flux rope would then drain from the system as it continued to expand \citep[][]{Mackay:2010}. For the event presented in this paper, the first obvious observation of material draining is seen to have initiated on the eastern side at $\sim$~01:30~UT. Then, after approximately 3 hours of sporadic and varied mass motions, the largest mass deposit is observed at $\sim$~04:30~UT by SDO/AIA to have propagated towards the eastern footpoint, as shown in Figure~\ref{fig:easttrack}. Interestingly, this large decrease in mass is shown in Figure~\ref{fig:kinematics} to have actually initiated at 04:00~UT ($\approx$~30 minutes prior) though this discrepancy is likely due to difficulties in isolating the initiation of the mass motion amongst all the additional, unrelated intensity variations shown in panels b and c of Figure~\ref{fig:easttrack}. We therefore conclude the large mass-unloading began at 04:00~UT. This rapidly reduced the mass contained within the filament, preceding the beginning of the `steep exponential phase' of the h--t profile of Figure~\ref{fig:kinematics}, and the splitting of the filament into two separate structures.
    
    According to the mass-unloading model described in \citet{Klimchuk:2001}, the presence of a sufficiently large mass within a non-potential system would allow this system to build free energy without a corresponding expansion, until the removal of the anchoring filament mass. Indeed if the mass-unloading was responsible for the change between the two exponential expansions, the ratio between gravitational forces supplied to the flux rope \textit{vs.} the forces acting down on the flux rope from above, must be on the order of or greater than one, \textit{i.e.,} the buoyant flux rope must have been able to overcome the restricting magnetic tension forces in order for the system to have accelerated. We can explore this through an order of magnitude estimation of the magnetic tension force, as outlined in the derivation presented in (Equation 6.2.18) of \citet{Aschwanden:2005}. The resulting ratio between the gravitational forces and magnetic forces is therefore,
\begin{equation}
     Ratio = \frac{\displaystyle (\Delta \rho) g}{\displaystyle \left(\frac{\displaystyle B^2}{\displaystyle \mu _0 r_c}\right)},
\end{equation}
where $\rho$ is plasma density, $g$ is acceleration due to gravity, $\mu _0$ is the permeability of free space, and $B$ and $r_c$ are the magnitude of the magnetic field and radius of curvature of the overlying loops, respectively, that supply the tension acting down on the flux rope.
    
	The value of mass unloaded is obtained from the STEREO-B/EUVI column density estimates of Section~\ref{S-density}. Using the assumption of a filament slab of dimensions 2~$\times$~10$^{8}$, 20~$\times$~10$^{8}$, and 200~$\times$~10$^{8}$~cm, the volume is estimated to be 8~$\times$~10$^{27}$~cm$^{3}$ and therefore the change in density is calculated to be 8.25~$\times$~10$^{-15}$~g~cm$^{-3}$. The magnitude of $B$ for the field overlying the flux rope is estimated by computing a potential field source surface (PFSS) model \corr{at 06:04~UT, at the start of the eruption} \citep[\textit{c.f.,}][]{Schrijver:2003}. \corr{The PFSS model of the coronal magnetic field is extrapolated using photospheric boundary conditions that are updated in six hour intervals. Though the extrapolation is carried out on a post-eruption photospheric magnetic field, this was deemed to more closely match the photospheric conditions at the time of eruption than those present in the magnetogram taken $\sim$~5 hours prior.} The radius of curvature for the overlying field is taken as a range, the lower bound of which is specified as the height the filament can be seen at by STEREO-B when the h--t profile changes from the shallow exponential to the steep exponential phase, therefore $r_c$~$\approx$~70 - 90~Mm, yielding a magnetic field strength $B$ for the \corr{apex of the potential} field overlying the flux rope of between 3 and 2.2~G. This produces a ratio between gravitational and magnetic tension forces of 1.8 to 4.1 respectively at 04:40~UT, the change from shallow to steep exponential phase. Therefore the height increase associated temporally with the mass-unloading shown in Figure~\ref{fig:kinematics} is interpreted as this expansion of the magnetic field due to the weakening of the anchoring force supplied by the large, dense filament present. In addition, the gravitational and kinetic energy of the system is calculated to have increased by 1.1~$\times$~10$^{28}$~ergs and 4.8~$\times$~10$^{26}$~ergs respectively due to this unloading of mass.
    
    As previously mentioned, the filament is seen to have split into two distinct structures in the lead up to its eruption, the dynamic portion that erupts and the restrained portion that remains. Studies have been carried out to better understand how a filament can split into two during their evolution; \citet{Gilbert:2001} describe a partial filament eruption and offer an explanation to how the magnetic environment, in their case a flux rope, can evolve and separate during an eruption. In the event that we have presented here, the identical types of observations are not available for study and so we are not able to test the applicability to their model. However, as both of our split filaments are assumed to involve flux ropes, it is reasonable to assume that this event shares some similarity to theirs. According to Figure~\ref{fig:splitting}, the separation can be seen to have occurred by $\approx$~05:26~UT, after the initiation of the large mass-unloading. As demonstrated above, it is interpreted that the mass-unloading was responsible for the change in the nature of the expansion of the filament from a shallow to steep exponential rise. It also appears that the mass-unloading was sourced at the apex of the dynamic portion of the filament, reducing the anchoring force of this region of the filament to the surface. With no noticeable mass-unloading having occurred within the region that becomes the restrained portion, it is reasonable to assume that this region did not experience the same reduced anchoring force and expansion that the dynamic portion underwent. Therefore, during the expansion of the dynamic portion, it is possible that the magnetic structure underwent some form of vertical reconnection due to the rise, as in the model presented by \citet{Gilbert:2001}, and permitted the splitting of the proposed flux rope into two. Interestingly, the material suspended in the restrained portion of the proposed flux rope, in the event presented here, is visibly perturbed during the eruption of the dynamic portion. This perturbed material is seen to then reform the restrained portion of the filament some hours after the eruption, suggesting that the magnetic structure of the restrained portion did not reconfigure to a significant degree during the nearby eruption and perturbation.
    
	Finally, both the monochromatic and polychromatic column density determination methods were applied to the filament in this investigation. The monochromatic method returns a lower estimate of the column density (which in itself is already a lower limit) as there is insufficient data to constrain filling factor and foreground emission fraction. Therefore the absolute values calculated are treated as estimations. However, the internal mass structure is well highlighted by examining the results of this method applied to the target at multiple times, and an evolution in mass can be estimated. While the polychromatic method was able to constrain filling factor and foreground emission, and hence give a more certain lower-limit on column density, it is not possible to infer a mass for the whole filament in this instance due to the uncertainty of the unattenuated radiation field behind the filament. The column density estimates of portions of the filament material derived using the polychromatic method, are found to be almost two orders of magnitude greater than for the monochromatic technique. This suggests that either this structure is far more dense than the STEREO-B data indicates, or the filament structure indeed resembled a slab topology, and SDO was simply observing a thicker structure from above than STEREO-B was from the side. A topological description of the slab structure from the magnetic dips in a flux rope may be found in the simulations carried out by authors such as \citet{Regnier:2004}, and \citet{Hillier:2013}. Nevertheless, if the density estimates of SDO are more representative of the filament, the ratio of gravitational forces to magnetic forces would increase by the corresponding orders of magnitude, indicating the mass-unloading played a significant role in the final evolution of the filament.%

\section{Conclusions} 
  \label{S-conclusions}


We have presented a multi-wavelength study of the pre-eruption period of the partial filament eruption on 11 December 2011 using data from two spacecraft. The multiple viewpoints have revealed the height-response of the filament due to material dynamics within; a separation otherwise unachievable from a single perspective. Four main stages of evolution were isolated: a stable phase (12:00 - $\sim$~13:00~UT), a rise phase (13:00 - 01:00~UT), a shallow exponential phase (01:00 - 04:40~UT), and a steep exponential phase (04:40 onwards). The rise phase temporally coincides with flux cancellation along the PIL below the filament of interest. Similarly, the nearby bipole emergence that was favourably oriented for reconnection with the magnetic arcade overlying and restraining the filament, the proposed flux rope becoming kink unstable, and the continued expansion of the filament through the solar atmosphere are seen to be contemporaneous, potentially highlighting the cancellation, emergence, and kinking as the triggers for the partial eruption \citep{Wang:1999,Torok:2005}. The proposed flux rope containing the filament is then seen to have become marginally unstable, demonstrated by a shallow exponential evolution in the height of the filament at 01:00~UT. Importantly, a large mass deposit beginning at 04:00~UT corresponding to a decrease in filament mass of 70$\%$ \citep[a larger percentage than those reported previously by \citet{Bi:2014} and][]{Fan:2017} yielded a lower-limit ratio range of 1.8 to 4.1 between gravitational forces and magnetic tension forces. The expansion of the dynamic portion of the filament is then seen to accelerate to a large exponential, we therefore conclude that the observed mass-unloading was responsible for the transition between the two exponential expansions of the filament.

\begin{acks}
The authors would like to thank the anonymous referee for their useful comments that significantly improved the quality and clarity of the manuscript. The authors wish to thank Magnus Woods, Alexander James, Francesco Zuccarello, and the attendees of the Flux Emergence Workshop 2017 for useful comments which improved the clarity of the paper, Pascal D{\'e}moulin and Gherardo Valori for insightful discussions and assistance with the PFSS modeling, and the efforts of the JHelioviewer team for their continued development of the JHelioviewer tool \citep{Mueller:2017}. JMJ thanks the STFC for support via funding given in his PhD Studentship. DML is an Early-Career Fellow, funded by the Leverhulme Trust. LvDG acknowledges funding under STFC consolidated grant number ST/N000722/1, Leverhulme Trust Research Project Grant 2014-051, and the Hungarian Research grant OTKA K-109276. JC is supported by an ESA Postdoctoral Research Fellowship at ESTEC. SDO is a mission of NASA's Living With a Star Program. STEREO is the third mission in NASA's Solar Terrestrial Probes program. The authors thank the SDO and STEREO teams for making their data publicly accessible.
\end{acks}




\bibliographystyle{spr-mp-sola}
\bibliography{bibliography} 

\begin{thebibliography}{67}
\ifx\bisbn     \undefined \def\bisbn  #1{ISBN #1}\fi
\ifx\binits    \undefined \def\binits#1{#1}\fi
\ifx\bauthor   \undefined \def\bauthor#1{#1}\fi
\ifx\batitle   \undefined \def\batitle#1{#1}\fi
\ifx\bjtitle   \undefined \def\bjtitle#1{\textit{#1}}\fi
\ifx\bvolume   \undefined \def\bvolume#1{\textbf{#1}}\fi
\ifx\byear     \undefined \def\byear#1{#1}\fi
\ifx\bissue    \undefined \def\bissue#1{#1}\fi
\ifx\bfpage    \undefined \def\bfpage#1{#1}\fi
\ifx\blpage    \undefined \def\blpage #1{#1}\fi
\ifx\burl      \undefined \def\burl#1{\textsf{#1}}\fi
\ifx\href      \undefined \def\href#1#2{\textsf{#2}}\fi
\ifx\betal     \undefined \def\betal{\textit{et al.}}\fi
\ifx\bctitle   \undefined \def\bctitle#1{#1}\fi
\ifx\beditor   \undefined \def\beditor#1{#1}\fi
\ifx\bbtitle   \undefined \def\bbtitle#1{\textit{#1}}\fi
\ifx\bedition  \undefined \def\bedition#1{#1}\fi
\ifx\bseriesno \undefined \def\bseriesno#1{\textbf{#1}}\fi
\ifx\blocation \undefined \def\blocation#1{#1}\fi
\ifx\bsertitle \undefined \def\bsertitle#1{\textit{#1}}\fi
\ifx\bsnm      \undefined \def\bsnm#1{#1}\fi
\ifx\bsuffix   \undefined \def\bsuffix#1{#1}\fi
\ifx\bparticle \undefined \def\bparticle#1{#1}\fi
\ifx\barticle  \undefined \def\barticle#1{}\fi
\ifx\binstitute  \undefined \def\binstitute#1{#1}\fi
\ifx\bpublisher  \undefined \def\bpublisher#1{#1}\fi
\ifx\doiurl    \undefined
  \def\doiurl#1{\href{http://dx.doi.org/#1}{\textsf{DOI}}}\fi
\ifx\arxivurl  \undefined
  \def\arxivurl#1{\href{http://arxiv.org/abs/#1}{\textsf{arXiv}}}\fi
\ifx\adsurl    \undefined
  \def\adsurl#1{\href{http://adsabs.harvard.edu/abs/#1}{\textsf{ADS}}}\fi
\ifx\botherref \undefined \def\botherref#1{}\fi
\ifx\url       \undefined \def\url#1{\textsf{#1}}\fi
\ifx\bchapter  \undefined \def\bchapter#1{}\fi
\ifx\bbook     \undefined \def\bbook#1{}\fi
\ifx\bcomment  \undefined \def\bcomment#1{#1}\fi
\ifx\oauthor   \undefined \def\oauthor#1{#1}\fi
\ifx\citeauthoryear \undefined\def \citeauthoryear#1{#1}\fi
\ifx\endbibitem\undefined \def\endbibitem{}\fi
\ifx\bconflocation  \undefined \def\bconflocation#1{#1} \fi

\bibitem[\protect\citeauthoryear{{Aschwanden}}{2005}]{Aschwanden:2005}
\begin{bbook}
\bauthor{\bsnm{{Aschwanden}}, \binits{M.J.}}:
\byear{2005},
\bbtitle{{Physics of the Solar Corona. An Introduction with Problems and
  Solutions (2nd edition)}}.
\adsurl{2005psci.book.....A}.
\end{bbook}
\endbibitem

\bibitem[\protect\citeauthoryear{{Aulanier} and
  {D\'emoulin}}{1998}]{Aulanier:1998}
\begin{barticle}
\bauthor{\bsnm{{Aulanier}}, \binits{G.}},
\bauthor{\bsnm{{D\'emoulin}}, \binits{P.}}:
\byear{1998},
\batitle{{3-D magnetic configurations supporting prominences. I. The natural
  presence of lateral feet}}.
\bjtitle{\aap}
\bvolume{329},
\bfpage{1125}.
\adsurl{1998A\%26A...329.1125A}.
\end{barticle}
\endbibitem

\bibitem[\protect\citeauthoryear{{Aulanier}
  \textit{et~al.}}{2010}]{Aulanier:2010}
\begin{barticle}
\bauthor{\bsnm{{Aulanier}}, \binits{G.}},
\bauthor{\bsnm{{T{\"o}r{\"o}k}}, \binits{T.}},
\bauthor{\bsnm{{D{\'e}moulin}}, \binits{P.}},
\bauthor{\bsnm{{DeLuca}}, \binits{E.E.}}:
\byear{2010},
\batitle{{Formation of Torus-Unstable Flux Ropes and Electric Currents in
  Erupting Sigmoids}}.
\bjtitle{\apj}
\bvolume{708},
\bfpage{314}.
\doiurl{10.1088/0004-637X/708/1/314}.
\adsurl{2010ApJ...708..314A}.
\end{barticle}
\endbibitem

\bibitem[\protect\citeauthoryear{{Bateman}}{1978}]{Bateman:1978}
\begin{bbook}
\bauthor{\bsnm{{Bateman}}, \binits{G.}}:
\byear{1978},
\bbtitle{{MHD instabilities}}.
\adsurl{1978mit..book.....B}.
\end{bbook}
\endbibitem

\bibitem[\protect\citeauthoryear{{Bi} \textit{et~al.}}{2014}]{Bi:2014}
\begin{barticle}
\bauthor{\bsnm{{Bi}}, \binits{Y.}},
\bauthor{\bsnm{{Jiang}}, \binits{Y.}},
\bauthor{\bsnm{{Yang}}, \binits{J.}},
\bauthor{\bsnm{{Hong}}, \binits{J.}},
\bauthor{\bsnm{{Li}}, \binits{H.}},
\bauthor{\bsnm{{Yang}}, \binits{D.}},
\bauthor{\bsnm{{Yang}}, \binits{B.}}:
\byear{2014},
\batitle{{Solar Filament Material Oscillations and Drainage before Eruption}}.
\bjtitle{\apj}
\bvolume{790},
\bfpage{100}.
\doiurl{10.1088/0004-637X/790/2/100}.
\adsurl{2014ApJ...790..100B}.
\end{barticle}
\endbibitem

\bibitem[\protect\citeauthoryear{{Brueckner}
  \textit{et~al.}}{1995}]{Brueckner:1995}
\begin{barticle}
\bauthor{\bsnm{{Brueckner}}, \binits{G.E.}},
\bauthor{\bsnm{{Howard}}, \binits{R.A.}},
\bauthor{\bsnm{{Koomen}}, \binits{M.J.}},
\bauthor{\bsnm{{Korendyke}}, \binits{C.M.}},
\bauthor{\bsnm{{Michels}}, \binits{D.J.}},
\bauthor{\bsnm{{Moses}}, \binits{J.D.}},
\bauthor{\bsnm{{Socker}}, \binits{D.G.}},
\bauthor{\bsnm{{Dere}}, \binits{K.P.}},
\bauthor{\bsnm{{Lamy}}, \binits{P.L.}},
\bauthor{\bsnm{{Llebaria}}, \binits{A.}},
\bauthor{\bsnm{{Bout}}, \binits{M.V.}},
\bauthor{\bsnm{{Schwenn}}, \binits{R.}},
\bauthor{\bsnm{{Simnett}}, \binits{G.M.}},
\bauthor{\bsnm{{Bedford}}, \binits{D.K.}},
\bauthor{\bsnm{{Eyles}}, \binits{C.J.}}:
\byear{1995},
\batitle{{The Large Angle Spectroscopic Coronagraph (LASCO)}}.
\bjtitle{\solphys}
\bvolume{162},
\bfpage{357}.
\doiurl{10.1007/BF00733434}.
\adsurl{1995SoPh..162..357B}.
\end{barticle}
\endbibitem

\bibitem[\protect\citeauthoryear{{Carlyle}
  \textit{et~al.}}{2014}]{Carlyle:2014a}
\begin{barticle}
\bauthor{\bsnm{{Carlyle}}, \binits{J.}},
\bauthor{\bsnm{{Williams}}, \binits{D.R.}},
\bauthor{\bsnm{{van Driel-Gesztelyi}}, \binits{L.}},
\bauthor{\bsnm{{Innes}}, \binits{D.}},
\bauthor{\bsnm{{Hillier}}, \binits{A.}},
\bauthor{\bsnm{{Matthews}}, \binits{S.}}:
\byear{2014},
\batitle{{Investigating the Dynamics and Density Evolution of Returning Plasma
  Blobs from the 2011 June 7 Eruption}}.
\bjtitle{\apj}
\bvolume{782},
\bfpage{87}.
\doiurl{10.1088/0004-637X/782/2/87}.
\adsurl{2014ApJ...782...87C}.
\end{barticle}
\endbibitem

\bibitem[\protect\citeauthoryear{{Chen} and {Shibata}}{2000}]{Chen:2000}
\begin{barticle}
\bauthor{\bsnm{{Chen}}, \binits{P.F.}},
\bauthor{\bsnm{{Shibata}}, \binits{K.}}:
\byear{2000},
\batitle{{An Emerging Flux Trigger Mechanism for Coronal Mass Ejections}}.
\bjtitle{\apj}
\bvolume{545},
\bfpage{524}.
\doiurl{10.1086/317803}.
\adsurl{2000ApJ...545..524C}.
\end{barticle}
\endbibitem

\bibitem[\protect\citeauthoryear{{Cheung} and {Isobe}}{2014}]{Cheung:2014}
\begin{barticle}
\bauthor{\bsnm{{Cheung}}, \binits{M.C.M.}},
\bauthor{\bsnm{{Isobe}}, \binits{H.}}:
\byear{2014},
\batitle{{Flux Emergence (Theory)}}.
\bjtitle{Living Reviews in Solar Physics}
\bvolume{11},
\bfpage{3}.
\doiurl{10.12942/lrsp-2014-3}.
\adsurl{2014LRSP...11....3C}.
\end{barticle}
\endbibitem

\bibitem[\protect\citeauthoryear{{D{\'e}moulin}}{1998}]{Demoulin:1998}
\begin{bchapter}
\bauthor{\bsnm{{D{\'e}moulin}}, \binits{P.}}:
\byear{1998},
\bctitle{{Magnetic Fields in Filaments (Review)}}.
In: \beditor{\bsnm{{Webb}}, \binits{D.F.}},
\beditor{\bsnm{{Schmieder}}, \binits{B.}},
\beditor{\bsnm{{Rust}}, \binits{D.M.}} (eds.)
\bbtitle{IAU Colloq. 167: New Perspectives on Solar Prominences},
\bsertitle{Astronomical Society of the Pacific Conference Series}
\bseriesno{150},
\bfpage{78}.
\adsurl{1998ASPC..150...78D}.
\end{bchapter}
\endbibitem

\bibitem[\protect\citeauthoryear{{Ding} and {Hu}}{2008}]{Ding:2008}
\begin{barticle}
\bauthor{\bsnm{{Ding}}, \binits{J.Y.}},
\bauthor{\bsnm{{Hu}}, \binits{Y.Q.}}:
\byear{2008},
\batitle{{Coronal Flux Rope Catastrophe Caused by Photospheric Flux
  Emergence}}.
\bjtitle{\apj}
\bvolume{674},
\bfpage{554}.
\doiurl{10.1086/524937}.
\adsurl{2008ApJ...674..554D}.
\end{barticle}
\endbibitem

\bibitem[\protect\citeauthoryear{{Domingo}, {Fleck}, and
  {Poland}}{1995}]{Domingo:1995}
\begin{barticle}
\bauthor{\bsnm{{Domingo}}, \binits{V.}},
\bauthor{\bsnm{{Fleck}}, \binits{B.}},
\bauthor{\bsnm{{Poland}}, \binits{A.I.}}:
\byear{1995},
\batitle{{The SOHO Mission: an Overview}}.
\bjtitle{\solphys}
\bvolume{162},
\bfpage{1}.
\doiurl{10.1007/BF00733425}.
\adsurl{1995SoPh..162....1D}.
\end{barticle}
\endbibitem

\bibitem[\protect\citeauthoryear{{Edl{\'e}n}}{1943}]{Edlen:1943}
\begin{barticle}
\bauthor{\bsnm{{Edl{\'e}n}}, \binits{B.}}:
\byear{1943},
\batitle{{Die Deutung der Emissionslinien im Spektrum der Sonnenkorona. Mit 6
  Abbildungen.}}
\bjtitle{\zap}
\bvolume{22},
\bfpage{30}.
\adsurl{1943ZA.....22...30E}.
\end{barticle}
\endbibitem

\bibitem[\protect\citeauthoryear{{Fan}}{2017}]{Fan:2017}
\begin{barticle}
\bauthor{\bsnm{{Fan}}, \binits{Y.}}:
\byear{2017},
\batitle{{MHD Simulations of the Eruption of Coronal Flux Ropes under Coronal
  Streamers}}.
\bjtitle{\apj}
\bvolume{844},
\bfpage{26}.
\doiurl{10.3847/1538-4357/aa7a56}.
\adsurl{2017ApJ...844...26F}.
\end{barticle}
\endbibitem

\bibitem[\protect\citeauthoryear{{Fan} and {Gibson}}{2007}]{Fan:2007}
\begin{barticle}
\bauthor{\bsnm{{Fan}}, \binits{Y.}},
\bauthor{\bsnm{{Gibson}}, \binits{S.E.}}:
\byear{2007},
\batitle{{Onset of Coronal Mass Ejections Due to Loss of Confinement of Coronal
  Flux Ropes}}.
\bjtitle{\apj}
\bvolume{668},
\bfpage{1232}.
\doiurl{10.1086/521335}.
\adsurl{2007ApJ...668.1232F}.
\end{barticle}
\endbibitem

\bibitem[\protect\citeauthoryear{{Feynman} and {Martin}}{1995}]{Feynman:1995}
\begin{barticle}
\bauthor{\bsnm{{Feynman}}, \binits{J.}},
\bauthor{\bsnm{{Martin}}, \binits{S.F.}}:
\byear{1995},
\batitle{{The initiation of coronal mass ejections by newly emerging magnetic
  flux}}.
\bjtitle{\jgr}
\bvolume{100},
\bfpage{3355}.
\doiurl{10.1029/94JA02591}.
\adsurl{1995JGR...100.3355F}.
\end{barticle}
\endbibitem

\bibitem[\protect\citeauthoryear{{Field}}{1965}]{Field:1965}
\begin{barticle}
\bauthor{\bsnm{{Field}}, \binits{G.B.}}:
\byear{1965},
\batitle{{Thermal Instability.}}
\bjtitle{\apj}
\bvolume{142},
\bfpage{531}.
\doiurl{10.1086/148317}.
\adsurl{1965ApJ...142..531F}.
\end{barticle}
\endbibitem

\bibitem[\protect\citeauthoryear{{Fleck}, {Couvidat}, and
  {Straus}}{2011}]{Fleck:2011}
\begin{barticle}
\bauthor{\bsnm{{Fleck}}, \binits{B.}},
\bauthor{\bsnm{{Couvidat}}, \binits{S.}},
\bauthor{\bsnm{{Straus}}, \binits{T.}}:
\byear{2011},
\batitle{{On the Formation Height of the SDO/HMI Fe 6173 {\AA} Doppler
  Signal}}.
\bjtitle{\solphys}
\bvolume{271},
\bfpage{27}.
\doiurl{10.1007/s11207-011-9783-9}.
\adsurl{2011SoPh..271...27F}.
\end{barticle}
\endbibitem

\bibitem[\protect\citeauthoryear{{Forbes}}{2000}]{Forbes:2000}
\begin{barticle}
\bauthor{\bsnm{{Forbes}}, \binits{T.G.}}:
\byear{2000},
\batitle{{A review on the genesis of coronal mass ejections}}.
\bjtitle{\jgr}
\bvolume{105},
\bfpage{23153}.
\doiurl{10.1029/2000JA000005}.
\adsurl{2000JGR...10523153F}.
\end{barticle}
\endbibitem

\bibitem[\protect\citeauthoryear{{Forland}
  \textit{et~al.}}{2013}]{Forland:2013}
\begin{barticle}
\bauthor{\bsnm{{Forland}}, \binits{B.C.}},
\bauthor{\bsnm{{Gibson}}, \binits{S.E.}},
\bauthor{\bsnm{{Dove}}, \binits{J.B.}},
\bauthor{\bsnm{{Rachmeler}}, \binits{L.A.}},
\bauthor{\bsnm{{Fan}}, \binits{Y.}}:
\byear{2013},
\batitle{{Coronal Cavity Survey: Morphological Clues to Eruptive Magnetic
  Topologies}}.
\bjtitle{\solphys}
\bvolume{288},
\bfpage{603}.
\doiurl{10.1007/s11207-013-0361-1}.
\adsurl{2013SoPh..288..603F}.
\end{barticle}
\endbibitem

\bibitem[\protect\citeauthoryear{{Gibson} \textit{et~al.}}{2006}]{Gibson:2006}
\begin{barticle}
\bauthor{\bsnm{{Gibson}}, \binits{S.E.}},
\bauthor{\bsnm{{Foster}}, \binits{D.}},
\bauthor{\bsnm{{Burkepile}}, \binits{J.}},
\bauthor{\bsnm{{de Toma}}, \binits{G.}},
\bauthor{\bsnm{{Stanger}}, \binits{A.}}:
\byear{2006},
\batitle{{The Calm before the Storm: The Link between Quiescent Cavities and
  Coronal Mass Ejections}}.
\bjtitle{\apj}
\bvolume{641},
\bfpage{590}.
\doiurl{10.1086/500446}.
\adsurl{2006ApJ...641..590G}.
\end{barticle}
\endbibitem

\bibitem[\protect\citeauthoryear{{Gilbert}, {Holzer}, and
  {Burkepile}}{2001}]{Gilbert:2001}
\begin{barticle}
\bauthor{\bsnm{{Gilbert}}, \binits{H.R.}},
\bauthor{\bsnm{{Holzer}}, \binits{T.E.}},
\bauthor{\bsnm{{Burkepile}}, \binits{J.T.}}:
\byear{2001},
\batitle{{Observational Interpretation of an Active Prominence on 1999 May 1}}.
\bjtitle{\apj}
\bvolume{549},
\bfpage{1221}.
\doiurl{10.1086/319444}.
\adsurl{2001ApJ...549.1221G}.
\end{barticle}
\endbibitem

\bibitem[\protect\citeauthoryear{{Green} \textit{et~al.}}{2007}]{Green:2007}
\begin{barticle}
\bauthor{\bsnm{{Green}}, \binits{L.M.}},
\bauthor{\bsnm{{Kliem}}, \binits{B.}},
\bauthor{\bsnm{{T{\"o}r{\"o}k}}, \binits{T.}},
\bauthor{\bsnm{{van Driel-Gesztelyi}}, \binits{L.}},
\bauthor{\bsnm{{Attrill}}, \binits{G.D.R.}}:
\byear{2007},
\batitle{{Transient Coronal Sigmoids and Rotating Erupting Flux Ropes}}.
\bjtitle{\solphys}
\bvolume{246},
\bfpage{365}.
\doiurl{10.1007/s11207-007-9061-z}.
\adsurl{2007SoPh..246..365G}.
\end{barticle}
\endbibitem

\bibitem[\protect\citeauthoryear{{Grevesse}, {Asplund}, and
  {Sauval}}{2007}]{Grevesse:2007}
\begin{barticle}
\bauthor{\bsnm{{Grevesse}}, \binits{N.}},
\bauthor{\bsnm{{Asplund}}, \binits{M.}},
\bauthor{\bsnm{{Sauval}}, \binits{A.J.}}:
\byear{2007},
\batitle{{The Solar Chemical Composition}}.
\bjtitle{\ssr}
\bvolume{130},
\bfpage{105}.
\doiurl{10.1007/s11214-007-9173-7}.
\adsurl{2007SSRv..130..105G}.
\end{barticle}
\endbibitem

\bibitem[\protect\citeauthoryear{{Gun{\'a}r} and {Mackay}}{2015}]{Gunar:2015}
\begin{barticle}
\bauthor{\bsnm{{Gun{\'a}r}}, \binits{S.}},
\bauthor{\bsnm{{Mackay}}, \binits{D.H.}}:
\byear{2015},
\batitle{{3D Whole-Prominence Fine Structure Modeling}}.
\bjtitle{\apj}
\bvolume{803},
\bfpage{64}.
\doiurl{10.1088/0004-637X/803/2/64}.
\adsurl{2015ApJ...803...64G}.
\end{barticle}
\endbibitem

\bibitem[\protect\citeauthoryear{{Gun{\'a}r}
  \textit{et~al.}}{2013}]{Gunar:2013}
\begin{barticle}
\bauthor{\bsnm{{Gun{\'a}r}}, \binits{S.}},
\bauthor{\bsnm{{Mackay}}, \binits{D.H.}},
\bauthor{\bsnm{{Anzer}}, \binits{U.}},
\bauthor{\bsnm{{Heinzel}}, \binits{P.}}:
\byear{2013},
\batitle{{Non-linear force-free magnetic dip models of quiescent prominence
  fine structures}}.
\bjtitle{\aap}
\bvolume{551},
\bfpage{A3}.
\doiurl{10.1051/0004-6361/201220597}.
\adsurl{2013A\%26A...551A...3G}.
\end{barticle}
\endbibitem

\bibitem[\protect\citeauthoryear{{Guo} \textit{et~al.}}{2017}]{Guo:2017}
\begin{barticle}
\bauthor{\bsnm{{Guo}}, \binits{Y.}},
\bauthor{\bsnm{{Pariat}}, \binits{E.}},
\bauthor{\bsnm{{Valori}}, \binits{G.}},
\bauthor{\bsnm{{Anfinogentov}}, \binits{S.}},
\bauthor{\bsnm{{Chen}}, \binits{F.}},
\bauthor{\bsnm{{Georgoulis}}, \binits{M.K.}},
\bauthor{\bsnm{{Liu}}, \binits{Y.}},
\bauthor{\bsnm{{Moraitis}}, \binits{K.}},
\bauthor{\bsnm{{Thalmann}}, \binits{J.K.}},
\bauthor{\bsnm{{Yang}}, \binits{S.}}:
\byear{2017},
\batitle{{Magnetic Helicity Estimations in Models and Observations of the Solar
  Magnetic Field. III. Twist Number Method}}.
\bjtitle{\apj}
\bvolume{840},
\bfpage{40}.
\doiurl{10.3847/1538-4357/aa6aa8}.
\adsurl{2017ApJ...840...40G}.
\end{barticle}
\endbibitem

\bibitem[\protect\citeauthoryear{{Heinzel}
  \textit{et~al.}}{2008}]{Heinzel:2008}
\begin{barticle}
\bauthor{\bsnm{{Heinzel}}, \binits{P.}},
\bauthor{\bsnm{{Schmieder}}, \binits{B.}},
\bauthor{\bsnm{{F{\'a}rn{\'{\i}}k}}, \binits{F.}},
\bauthor{\bsnm{{Schwartz}}, \binits{P.}},
\bauthor{\bsnm{{Labrosse}}, \binits{N.}},
\bauthor{\bsnm{{Kotr{\v c}}}, \binits{P.}},
\bauthor{\bsnm{{Anzer}}, \binits{U.}},
\bauthor{\bsnm{{Molodij}}, \binits{G.}},
\bauthor{\bsnm{{Berlicki}}, \binits{A.}},
\bauthor{\bsnm{{DeLuca}}, \binits{E.E.}},
\bauthor{\bsnm{{Golub}}, \binits{L.}},
\bauthor{\bsnm{{Watanabe}}, \binits{T.}},
\bauthor{\bsnm{{Berger}}, \binits{T.}}:
\byear{2008},
\batitle{{Hinode, TRACE, SOHO, and Ground-based Observations of a Quiescent
  Prominence}}.
\bjtitle{\apj}
\bvolume{686},
\bfpage{1383}.
\doiurl{10.1086/591018}.
\adsurl{2008ApJ...686.1383H}.
\end{barticle}
\endbibitem

\bibitem[\protect\citeauthoryear{{Hillier} and {van
  Ballegooijen}}{2013}]{Hillier:2013}
\begin{barticle}
\bauthor{\bsnm{{Hillier}}, \binits{A.}},
\bauthor{\bsnm{{van Ballegooijen}}, \binits{A.}}:
\byear{2013},
\batitle{{On the Support of Solar Prominence Material by the Dips of a Coronal
  Flux Tube}}.
\bjtitle{\apj}
\bvolume{766},
\bfpage{126}.
\doiurl{10.1088/0004-637X/766/2/126}.
\adsurl{2013ApJ...766..126H}.
\end{barticle}
\endbibitem

\bibitem[\protect\citeauthoryear{{Hoeksema}
  \textit{et~al.}}{2014}]{Hoeksema:2014}
\begin{barticle}
\bauthor{\bsnm{{Hoeksema}}, \binits{J.T.}},
\bauthor{\bsnm{{Liu}}, \binits{Y.}},
\bauthor{\bsnm{{Hayashi}}, \binits{K.}},
\bauthor{\bsnm{{Sun}}, \binits{X.}},
\bauthor{\bsnm{{Schou}}, \binits{J.}},
\bauthor{\bsnm{{Couvidat}}, \binits{S.}},
\bauthor{\bsnm{{Norton}}, \binits{A.}},
\bauthor{\bsnm{{Bobra}}, \binits{M.}},
\bauthor{\bsnm{{Centeno}}, \binits{R.}},
\bauthor{\bsnm{{Leka}}, \binits{K.D.}},
\bauthor{\bsnm{{Barnes}}, \binits{G.}},
\bauthor{\bsnm{{Turmon}}, \binits{M.}}:
\byear{2014},
\batitle{{The Helioseismic and Magnetic Imager (HMI) Vector Magnetic Field
  Pipeline: Overview and Performance}}.
\bjtitle{\solphys}
\bvolume{289},
\bfpage{3483}.
\doiurl{10.1007/s11207-014-0516-8}.
\adsurl{2014SoPh..289.3483H}.
\end{barticle}
\endbibitem

\bibitem[\protect\citeauthoryear{{Hood} and {Priest}}{1981}]{Hood:1981}
\begin{barticle}
\bauthor{\bsnm{{Hood}}, \binits{A.W.}},
\bauthor{\bsnm{{Priest}}, \binits{E.R.}}:
\byear{1981},
\batitle{{Critical conditions for magnetic instabilities in force-free coronal
  loops}}.
\bjtitle{Geophysical and Astrophysical Fluid Dynamics}
\bvolume{17},
\bfpage{297}.
\doiurl{10.1080/03091928108243687}.
\adsurl{1981GApFD..17..297H}.
\end{barticle}
\endbibitem

\bibitem[\protect\citeauthoryear{{James} \textit{et~al.}}{2017}]{James:2017}
\begin{barticle}
\bauthor{\bsnm{{James}}, \binits{A.W.}},
\bauthor{\bsnm{{Green}}, \binits{L.M.}},
\bauthor{\bsnm{{Palmerio}}, \binits{E.}},
\bauthor{\bsnm{{Valori}}, \binits{G.}},
\bauthor{\bsnm{{Reid}}, \binits{H.A.S.}},
\bauthor{\bsnm{{Baker}}, \binits{D.}},
\bauthor{\bsnm{{Brooks}}, \binits{D.H.}},
\bauthor{\bsnm{{van Driel-Gesztelyi}}, \binits{L.}},
\bauthor{\bsnm{{Kilpua}}, \binits{E.K.J.}}:
\byear{2017},
\batitle{{On-Disc Observations of Flux Rope Formation Prior to Its Eruption}}.
\bjtitle{\solphys}
\bvolume{292},
\bfpage{71}.
\doiurl{10.1007/s11207-017-1093-4}.
\adsurl{2017SoPh..292...71J}.
\end{barticle}
\endbibitem

\bibitem[\protect\citeauthoryear{{Jiang} \textit{et~al.}}{2014}]{Jiang:2014}
\begin{barticle}
\bauthor{\bsnm{{Jiang}}, \binits{C.}},
\bauthor{\bsnm{{Wu}}, \binits{S.T.}},
\bauthor{\bsnm{{Feng}}, \binits{X.}},
\bauthor{\bsnm{{Hu}}, \binits{Q.}}:
\byear{2014},
\batitle{{Nonlinear Force-free Field Extrapolation of a Coronal Magnetic Flux
  Rope Supporting a Large-scale Solar Filament from a Photospheric Vector
  Magnetogram}}.
\bjtitle{\apjl}
\bvolume{786},
\bfpage{L16}.
\doiurl{10.1088/2041-8205/786/2/L16}.
\adsurl{2014ApJ...786L..16J}.
\end{barticle}
\endbibitem

\bibitem[\protect\citeauthoryear{{Kaiser} \textit{et~al.}}{2008}]{Kaiser:2008}
\begin{barticle}
\bauthor{\bsnm{{Kaiser}}, \binits{M.L.}},
\bauthor{\bsnm{{Kucera}}, \binits{T.A.}},
\bauthor{\bsnm{{Davila}}, \binits{J.M.}},
\bauthor{\bsnm{{St.~Cyr}}, \binits{O.C.}},
\bauthor{\bsnm{{Guhathakurta}}, \binits{M.}},
\bauthor{\bsnm{{Christian}}, \binits{E.}}:
\byear{2008},
\batitle{{The STEREO Mission: An Introduction}}.
\bjtitle{\ssr}
\bvolume{136},
\bfpage{5}.
\doiurl{10.1007/s11214-007-9277-0}.
\adsurl{2008SSRv..136....5K}.
\end{barticle}
\endbibitem

\bibitem[\protect\citeauthoryear{{Kliem} and
  {T{\"o}r{\"o}k}}{2006}]{Kliem:2006}
\begin{barticle}
\bauthor{\bsnm{{Kliem}}, \binits{B.}},
\bauthor{\bsnm{{T{\"o}r{\"o}k}}, \binits{T.}}:
\byear{2006},
\batitle{{Torus Instability}}.
\bjtitle{Physical Review Letters}
\bvolume{96}(\bissue{25}),
\bfpage{255002}.
\doiurl{10.1103/PhysRevLett.96.255002}.
\adsurl{2006PhRvL..96y5002K}.
\end{barticle}
\endbibitem

\bibitem[\protect\citeauthoryear{{Klimchuk}}{2001}]{Klimchuk:2001}
\begin{barticle}
\bauthor{\bsnm{{Klimchuk}}, \binits{J.A.}}:
\byear{2001},
\batitle{{Theory of Coronal Mass Ejections}}.
\bjtitle{Washington DC American Geophysical Union Geophysical Monograph Series}
\bvolume{125}.
\doiurl{10.1029/GM125p0143}.
\adsurl{2001GMS...125..143K}.
\end{barticle}
\endbibitem

\bibitem[\protect\citeauthoryear{{Landi} and {Reale}}{2013}]{Landi:2013}
\begin{barticle}
\bauthor{\bsnm{{Landi}}, \binits{E.}},
\bauthor{\bsnm{{Reale}}, \binits{F.}}:
\byear{2013},
\batitle{{Prominence Plasma Diagnostics through Extreme-ultraviolet
  Absorption}}.
\bjtitle{\apj}
\bvolume{772},
\bfpage{71}.
\doiurl{10.1088/0004-637X/772/1/71}.
\adsurl{2013ApJ...772...71L}.
\end{barticle}
\endbibitem

\bibitem[\protect\citeauthoryear{{Lemen} \textit{et~al.}}{2012}]{Lemen:2012}
\begin{barticle}
\bauthor{\bsnm{{Lemen}}, \binits{J.R.}},
\bauthor{\bsnm{{Title}}, \binits{A.M.}},
\bauthor{\bsnm{{Akin}}, \binits{D.J.}},
\bauthor{\bsnm{{Boerner}}, \binits{P.F.}},
\bauthor{\bsnm{{Chou}}, \binits{C.}},
\bauthor{\bsnm{{Drake}}, \binits{J.F.}}:
\byear{2012},
\batitle{{The Atmospheric Imaging Assembly (AIA) on the Solar Dynamics
  Observatory (SDO)}}.
\bjtitle{\solphys}
\bvolume{275},
\bfpage{17}.
\doiurl{10.1007/s11207-011-9776-8}.
\adsurl{2012SoPh..275...17L}.
\end{barticle}
\endbibitem

\bibitem[\protect\citeauthoryear{{Lionello}
  \textit{et~al.}}{2002}]{Lionello:2002}
\begin{barticle}
\bauthor{\bsnm{{Lionello}}, \binits{R.}},
\bauthor{\bsnm{{Miki{\'c}}}, \binits{Z.}},
\bauthor{\bsnm{{Linker}}, \binits{J.A.}},
\bauthor{\bsnm{{Amari}}, \binits{T.}}:
\byear{2002},
\batitle{{Magnetic Field Topology in Prominences}}.
\bjtitle{\apj}
\bvolume{581},
\bfpage{718}.
\doiurl{10.1086/344222}.
\adsurl{2002ApJ...581..718L}.
\end{barticle}
\endbibitem

\bibitem[\protect\citeauthoryear{{Loboda} and {Bogachev}}{2015}]{Loboda:2015}
\begin{barticle}
\bauthor{\bsnm{{Loboda}}, \binits{I.P.}},
\bauthor{\bsnm{{Bogachev}}, \binits{S.A.}}:
\byear{2015},
\batitle{{Quiescent and Eruptive Prominences at Solar Minimum: A Statistical
  Study via an Automated Tracking System}}.
\bjtitle{\solphys}
\bvolume{290},
\bfpage{1963}.
\doiurl{10.1007/s11207-015-0735-7}.
\adsurl{2015SoPh..290.1963L}.
\end{barticle}
\endbibitem

\bibitem[\protect\citeauthoryear{{Low}}{1996}]{Low:1996}
\begin{barticle}
\bauthor{\bsnm{{Low}}, \binits{B.C.}}:
\byear{1996},
\batitle{{Solar Activity and the Corona}}.
\bjtitle{\solphys}
\bvolume{167},
\bfpage{217}.
\doiurl{10.1007/BF00146338}.
\adsurl{1996SoPh..167..217L}.
\end{barticle}
\endbibitem

\bibitem[\protect\citeauthoryear{{Mackay}, {Gaizauskas}, and
  {Yeates}}{2008}]{Mackay:2008}
\begin{barticle}
\bauthor{\bsnm{{Mackay}}, \binits{D.H.}},
\bauthor{\bsnm{{Gaizauskas}}, \binits{V.}},
\bauthor{\bsnm{{Yeates}}, \binits{A.R.}}:
\byear{2008},
\batitle{{Where Do Solar Filaments Form?: Consequences for Theoretical
  Models}}.
\bjtitle{\solphys}
\bvolume{248},
\bfpage{51}.
\doiurl{10.1007/s11207-008-9127-6}.
\adsurl{2008SoPh..248...51M}.
\end{barticle}
\endbibitem

\bibitem[\protect\citeauthoryear{{Mackay} \textit{et~al.}}{2010}]{Mackay:2010}
\begin{barticle}
\bauthor{\bsnm{{Mackay}}, \binits{D.H.}},
\bauthor{\bsnm{{Karpen}}, \binits{J.T.}},
\bauthor{\bsnm{{Ballester}}, \binits{J.L.}},
\bauthor{\bsnm{{Schmieder}}, \binits{B.}},
\bauthor{\bsnm{{Aulanier}}, \binits{G.}}:
\byear{2010},
\batitle{{Physics of Solar Prominences: II---Magnetic Structure and Dynamics}}.
\bjtitle{\ssr}
\bvolume{151},
\bfpage{333}.
\doiurl{10.1007/s11214-010-9628-0}.
\adsurl{2010SSRv..151..333M}.
\end{barticle}
\endbibitem

\bibitem[\protect\citeauthoryear{{McIntosh}
  \textit{et~al.}}{2014}]{McIntosh:2014}
\begin{barticle}
\bauthor{\bsnm{{McIntosh}}, \binits{S.W.}},
\bauthor{\bsnm{{Wang}}, \binits{X.}},
\bauthor{\bsnm{{Leamon}}, \binits{R.J.}},
\bauthor{\bsnm{{Davey}}, \binits{A.R.}},
\bauthor{\bsnm{{Howe}}, \binits{R.}},
\bauthor{\bsnm{{Krista}}, \binits{L.D.}},
\bauthor{\bsnm{{Malanushenko}}, \binits{A.V.}},
\bauthor{\bsnm{{Markel}}, \binits{R.S.}},
\bauthor{\bsnm{{Cirtain}}, \binits{J.W.}},
\bauthor{\bsnm{{Gurman}}, \binits{J.B.}},
\bauthor{\bsnm{{Pesnell}}, \binits{W.D.}},
\bauthor{\bsnm{{Thompson}}, \binits{M.J.}}:
\byear{2014},
\batitle{{Deciphering Solar Magnetic Activity. I. On the Relationship between
  the Sunspot Cycle and the Evolution of Small Magnetic Features}}.
\bjtitle{\apj}
\bvolume{792},
\bfpage{12}.
\doiurl{10.1088/0004-637X/792/1/12}.
\adsurl{2014ApJ...792...12M}.
\end{barticle}
\endbibitem

\bibitem[\protect\citeauthoryear{{Mei} \textit{et~al.}}{2017}]{Mei:2017}
\begin{barticle}
\bauthor{\bsnm{{Mei}}, \binits{Z.X.}},
\bauthor{\bsnm{{Keppens}}, \binits{R.}},
\bauthor{\bsnm{{Roussev}}, \binits{I.I.}},
\bauthor{\bsnm{{Lin}}, \binits{J.}}:
\byear{2017},
\batitle{{Magnetic reconnection during eruptive magnetic flux ropes}}.
\bjtitle{\aap}
\bvolume{604},
\bfpage{L7}.
\doiurl{10.1051/0004-6361/201731146}.
\adsurl{2017A\%26A...604L...7M}.
\end{barticle}
\endbibitem

\bibitem[\protect\citeauthoryear{{Moore} \textit{et~al.}}{2001}]{Moore:2001}
\begin{barticle}
\bauthor{\bsnm{{Moore}}, \binits{R.L.}},
\bauthor{\bsnm{{Sterling}}, \binits{A.C.}},
\bauthor{\bsnm{{Hudson}}, \binits{H.S.}},
\bauthor{\bsnm{{Lemen}}, \binits{J.R.}}:
\byear{2001},
\batitle{{Onset of the Magnetic Explosion in Solar Flares and Coronal Mass
  Ejections}}.
\bjtitle{\apj}
\bvolume{552},
\bfpage{833}.
\doiurl{10.1086/320559}.
\adsurl{2001ApJ...552..833M}.
\end{barticle}
\endbibitem

\bibitem[\protect\citeauthoryear{{Morgan} and
  {Druckm{\"u}ller}}{2014}]{Morgan:2014}
\begin{barticle}
\bauthor{\bsnm{{Morgan}}, \binits{H.}},
\bauthor{\bsnm{{Druckm{\"u}ller}}, \binits{M.}}:
\byear{2014},
\batitle{{Multi-Scale Gaussian Normalization for Solar Image Processing}}.
\bjtitle{\solphys}
\bvolume{289},
\bfpage{2945}.
\doiurl{10.1007/s11207-014-0523-9}.
\adsurl{2014SoPh..289.2945M}.
\end{barticle}
\endbibitem

\bibitem[\protect\citeauthoryear{{Mueller}
  \textit{et~al.}}{2017}]{Mueller:2017}
\begin{botherref}
\oauthor{\bsnm{{Mueller}}, \binits{D.}},
\oauthor{\bsnm{{Nicula}}, \binits{B.}},
\oauthor{\bsnm{{Felix}}, \binits{S.}},
\oauthor{\bsnm{{Verstringe}}, \binits{F.}},
\oauthor{\bsnm{{Bourgoignie}}, \binits{B.}},
\oauthor{\bsnm{{Csillaghy}}, \binits{A.}},
\oauthor{\bsnm{{Berghmans}}, \binits{D.}},
\oauthor{\bsnm{{Jiggens}}, \binits{P.}},
\oauthor{\bsnm{{Garcia-Ortiz}}, \binits{J.P.}},
\oauthor{\bsnm{{Ireland}}, \binits{J.}},
\oauthor{\bsnm{{Zahniy}}, \binits{S.}},
\oauthor{\bsnm{{Fleck}}, \binits{B.}}:
2017,
{JHelioviewer - Time-dependent 3D visualisation of solar and heliospheric
  data}.
\textit{ArXiv e-prints}.
\adsurl{2017arXiv170507628M}.
\end{botherref}
\endbibitem

\bibitem[\protect\citeauthoryear{{Palacios}
  \textit{et~al.}}{2015}]{Palacios:2015}
\begin{barticle}
\bauthor{\bsnm{{Palacios}}, \binits{J.}},
\bauthor{\bsnm{{Cid}}, \binits{C.}},
\bauthor{\bsnm{{Guerrero}}, \binits{A.}},
\bauthor{\bsnm{{Saiz}}, \binits{E.}},
\bauthor{\bsnm{{Cerrato}}, \binits{Y.}}:
\byear{2015},
\batitle{{Supergranular-scale magnetic flux emergence beneath an unstable
  filament}}.
\bjtitle{\aap}
\bvolume{583},
\bfpage{A47}.
\doiurl{10.1051/0004-6361/201323284}.
\adsurl{2015A\%26A...583A..47P}.
\end{barticle}
\endbibitem

\bibitem[\protect\citeauthoryear{{Parenti}}{2014}]{Parenti:2014}
\begin{barticle}
\bauthor{\bsnm{{Parenti}}, \binits{S.}}:
\byear{2014},
\batitle{{Solar Prominences: Observations}}.
\bjtitle{Living Reviews in Solar Physics}
\bvolume{11},
\bfpage{1}.
\doiurl{10.12942/lrsp-2014-1}.
\adsurl{2014LRSP...11....1P}.
\end{barticle}
\endbibitem

\bibitem[\protect\citeauthoryear{{Pesnell}, {Thompson}, and
  {Chamberlin}}{2012}]{Pesnell:2012}
\begin{barticle}
\bauthor{\bsnm{{Pesnell}}, \binits{W.D.}},
\bauthor{\bsnm{{Thompson}}, \binits{B.J.}},
\bauthor{\bsnm{{Chamberlin}}, \binits{P.C.}}:
\byear{2012},
\batitle{{The Solar Dynamics Observatory (SDO)}}.
\bjtitle{\solphys}
\bvolume{275},
\bfpage{3}.
\doiurl{10.1007/s11207-011-9841-3}.
\adsurl{2012SoPh..275....3P}.
\end{barticle}
\endbibitem

\bibitem[\protect\citeauthoryear{{Polito} \textit{et~al.}}{2017}]{Polito:2017}
\begin{barticle}
\bauthor{\bsnm{{Polito}}, \binits{V.}},
\bauthor{\bsnm{{Del Zanna}}, \binits{G.}},
\bauthor{\bsnm{{Valori}}, \binits{G.}},
\bauthor{\bsnm{{Pariat}}, \binits{E.}},
\bauthor{\bsnm{{Mason}}, \binits{H.E.}},
\bauthor{\bsnm{{Dud{\'{\i}}k}}, \binits{J.}},
\bauthor{\bsnm{{Janvier}}, \binits{M.}}:
\byear{2017},
\batitle{{Analysis and modelling of recurrent solar flares observed with
  Hinode/EIS on March 9, 2012}}.
\bjtitle{\aap}
\bvolume{601},
\bfpage{A39}.
\doiurl{10.1051/0004-6361/201629703}.
\adsurl{2017A\%26A...601A..39P}.
\end{barticle}
\endbibitem

\bibitem[\protect\citeauthoryear{{R{\'e}gnier} and
  {Amari}}{2004}]{Regnier:2004}
\begin{barticle}
\bauthor{\bsnm{{R{\'e}gnier}}, \binits{S.}},
\bauthor{\bsnm{{Amari}}, \binits{T.}}:
\byear{2004},
\batitle{{3D magnetic configuration of the H{$\alpha$} filament and X-ray
  sigmoid in NOAA AR 8151}}.
\bjtitle{\aap}
\bvolume{425},
\bfpage{345}.
\doiurl{10.1051/0004-6361:20034383}.
\adsurl{2004A\%26A...425..345R}.
\end{barticle}
\endbibitem

\bibitem[\protect\citeauthoryear{{Roussev}
  \textit{et~al.}}{2003}]{Roussev:2003}
\begin{barticle}
\bauthor{\bsnm{{Roussev}}, \binits{I.I.}},
\bauthor{\bsnm{{Forbes}}, \binits{T.G.}},
\bauthor{\bsnm{{Gombosi}}, \binits{T.I.}},
\bauthor{\bsnm{{Sokolov}}, \binits{I.V.}},
\bauthor{\bsnm{{DeZeeuw}}, \binits{D.L.}},
\bauthor{\bsnm{{Birn}}, \binits{J.}}:
\byear{2003},
\batitle{{A Three-dimensional Flux Rope Model for Coronal Mass Ejections Based
  on a Loss of Equilibrium}}.
\bjtitle{\apjl}
\bvolume{588},
\bfpage{L45}.
\doiurl{10.1086/375442}.
\adsurl{2003ApJ...588L..45R}.
\end{barticle}
\endbibitem

\bibitem[\protect\citeauthoryear{{Schrijver} and {De
  Rosa}}{2003}]{Schrijver:2003}
\begin{barticle}
\bauthor{\bsnm{{Schrijver}}, \binits{C.J.}},
\bauthor{\bsnm{{De Rosa}}, \binits{M.L.}}:
\byear{2003},
\batitle{{Photospheric and heliospheric magnetic fields}}.
\bjtitle{\solphys}
\bvolume{212},
\bfpage{165}.
\doiurl{10.1023/A:1022908504100}.
\adsurl{2003SoPh..212..165S}.
\end{barticle}
\endbibitem

\bibitem[\protect\citeauthoryear{{Schrijver}
  \textit{et~al.}}{2008}]{Schrijver:2008}
\begin{barticle}
\bauthor{\bsnm{{Schrijver}}, \binits{C.J.}},
\bauthor{\bsnm{{Elmore}}, \binits{C.}},
\bauthor{\bsnm{{Kliem}}, \binits{B.}},
\bauthor{\bsnm{{T{\"o}r{\"o}k}}, \binits{T.}},
\bauthor{\bsnm{{Title}}, \binits{A.M.}}:
\byear{2008},
\batitle{{Observations and Modeling of the Early Acceleration Phase of Erupting
  Filaments Involved in Coronal Mass Ejections}}.
\bjtitle{\apj}
\bvolume{674},
\bfpage{586}.
\doiurl{10.1086/524294}.
\adsurl{2008ApJ...674..586S}.
\end{barticle}
\endbibitem

\bibitem[\protect\citeauthoryear{{Seaton} \textit{et~al.}}{2011}]{Seaton:2011}
\begin{barticle}
\bauthor{\bsnm{{Seaton}}, \binits{D.B.}},
\bauthor{\bsnm{{Mierla}}, \binits{M.}},
\bauthor{\bsnm{{Berghmans}}, \binits{D.}},
\bauthor{\bsnm{{Zhukov}}, \binits{A.N.}},
\bauthor{\bsnm{{Dolla}}, \binits{L.}}:
\byear{2011},
\batitle{{SWAP-SECCHI Observations of a Mass-loading Type Solar Eruption}}.
\bjtitle{\apjl}
\bvolume{727},
\bfpage{L10}.
\doiurl{10.1088/2041-8205/727/1/L10}.
\adsurl{2011ApJ...727L..10S}.
\end{barticle}
\endbibitem

\bibitem[\protect\citeauthoryear{{Su} \textit{et~al.}}{2011}]{Su:2011}
\begin{barticle}
\bauthor{\bsnm{{Su}}, \binits{Y.}},
\bauthor{\bsnm{{Surges}}, \binits{V.}},
\bauthor{\bsnm{{van Ballegooijen}}, \binits{A.}},
\bauthor{\bsnm{{DeLuca}}, \binits{E.}},
\bauthor{\bsnm{{Golub}}, \binits{L.}}:
\byear{2011},
\batitle{{Observations and Magnetic Field Modeling of the Flare/coronal Mass
  Ejection Event on 2010 April 8}}.
\bjtitle{\apj}
\bvolume{734},
\bfpage{53}.
\doiurl{10.1088/0004-637X/734/1/53}.
\adsurl{2011ApJ...734...53S}.
\end{barticle}
\endbibitem

\bibitem[\protect\citeauthoryear{{Thompson}
  \textit{et~al.}}{2000}]{Thompson:2000}
\begin{barticle}
\bauthor{\bsnm{{Thompson}}, \binits{B.J.}},
\bauthor{\bsnm{{Cliver}}, \binits{E.W.}},
\bauthor{\bsnm{{Nitta}}, \binits{N.}},
\bauthor{\bsnm{{Delann{\'e}e}}, \binits{C.}},
\bauthor{\bsnm{{Delaboudini{\`e}re}}, \binits{J.-P.}}:
\byear{2000},
\batitle{{Coronal dimmings and energetic CMEs in April-May 1998}}.
\bjtitle{\grl}
\bvolume{27},
\bfpage{1431}.
\doiurl{10.1029/1999GL003668}.
\adsurl{2000GeoRL..27.1431T}.
\end{barticle}
\endbibitem

\bibitem[\protect\citeauthoryear{{T{\"o}r{\"o}k} and
  {Kliem}}{2005}]{Torok:2005}
\begin{barticle}
\bauthor{\bsnm{{T{\"o}r{\"o}k}}, \binits{T.}},
\bauthor{\bsnm{{Kliem}}, \binits{B.}}:
\byear{2005},
\batitle{{Confined and Ejective Eruptions of Kink-unstable Flux Ropes}}.
\bjtitle{\apjl}
\bvolume{630},
\bfpage{L97}.
\doiurl{10.1086/462412}.
\adsurl{2005ApJ...630L..97T}.
\end{barticle}
\endbibitem

\bibitem[\protect\citeauthoryear{{T{\"o}r{\"o}k}
  \textit{et~al.}}{2011}]{Torok:2011}
\begin{barticle}
\bauthor{\bsnm{{T{\"o}r{\"o}k}}, \binits{T.}},
\bauthor{\bsnm{{Panasenco}}, \binits{O.}},
\bauthor{\bsnm{{Titov}}, \binits{V.S.}},
\bauthor{\bsnm{{Miki{\'c}}}, \binits{Z.}},
\bauthor{\bsnm{{Reeves}}, \binits{K.K.}},
\bauthor{\bsnm{{Velli}}, \binits{M.}},
\bauthor{\bsnm{{Linker}}, \binits{J.A.}},
\bauthor{\bsnm{{De Toma}}, \binits{G.}}:
\byear{2011},
\batitle{{A Model for Magnetically Coupled Sympathetic Eruptions}}.
\bjtitle{\apjl}
\bvolume{739},
\bfpage{L63}.
\doiurl{10.1088/2041-8205/739/2/L63}.
\adsurl{2011ApJ...739L..63T}.
\end{barticle}
\endbibitem

\bibitem[\protect\citeauthoryear{{van Ballegooijen} and
  {Martens}}{1989}]{vanballegooijen:1989}
\begin{barticle}
\bauthor{\bsnm{{van Ballegooijen}}, \binits{A.A.}},
\bauthor{\bsnm{{Martens}}, \binits{P.C.H.}}:
\byear{1989},
\batitle{{Formation and eruption of solar prominences}}.
\bjtitle{\apj}
\bvolume{343},
\bfpage{971}.
\doiurl{10.1086/167766}.
\adsurl{1989ApJ...343..971V}.
\end{barticle}
\endbibitem

\bibitem[\protect\citeauthoryear{{Wang} and {Sheeley}}{1999}]{Wang:1999}
\begin{barticle}
\bauthor{\bsnm{{Wang}}, \binits{Y.-M.}},
\bauthor{\bsnm{{Sheeley}}, \binits{N.R.} \bsuffix{Jr.}}:
\byear{1999},
\batitle{{Filament Eruptions near Emerging Bipoles}}.
\bjtitle{\apjl}
\bvolume{510},
\bfpage{L157}.
\doiurl{10.1086/311815}.
\adsurl{1999ApJ...510L.157W}.
\end{barticle}
\endbibitem

\bibitem[\protect\citeauthoryear{{Webb} and {Howard}}{2012}]{Webb:2012}
\begin{barticle}
\bauthor{\bsnm{{Webb}}, \binits{D.F.}},
\bauthor{\bsnm{{Howard}}, \binits{T.A.}}:
\byear{2012},
\batitle{{Coronal Mass Ejections: Observations}}.
\bjtitle{Living Reviews in Solar Physics}
\bvolume{9},
\bfpage{3}.
\doiurl{10.12942/lrsp-2012-3}.
\adsurl{2012LRSP....9....3W}.
\end{barticle}
\endbibitem

\bibitem[\protect\citeauthoryear{{Williams}, {Baker}, and {van
  Driel-Gesztelyi}}{2013}]{Williams:2013}
\begin{barticle}
\bauthor{\bsnm{{Williams}}, \binits{D.R.}},
\bauthor{\bsnm{{Baker}}, \binits{D.}},
\bauthor{\bsnm{{van Driel-Gesztelyi}}, \binits{L.}}:
\byear{2013},
\batitle{{Mass Estimates of Rapidly Moving Prominence Material from
  High-cadence EUV Images}}.
\bjtitle{\apj}
\bvolume{764},
\bfpage{165}.
\doiurl{10.1088/0004-637X/764/2/165}.
\adsurl{2013ApJ...764..165W}.
\end{barticle}
\endbibitem

\bibitem[\protect\citeauthoryear{{Williams}
  \textit{et~al.}}{2005}]{Williams:2005}
\begin{barticle}
\bauthor{\bsnm{{Williams}}, \binits{D.R.}},
\bauthor{\bsnm{{T{\"o}r{\"o}k}}, \binits{T.}},
\bauthor{\bsnm{{D{\'e}moulin}}, \binits{P.}},
\bauthor{\bsnm{{van Driel-Gesztelyi}}, \binits{L.}},
\bauthor{\bsnm{{Kliem}}, \binits{B.}}:
\byear{2005},
\batitle{{Eruption of a Kink-unstable Filament in NOAA Active Region 10696}}.
\bjtitle{\apjl}
\bvolume{628},
\bfpage{L163}.
\doiurl{10.1086/432910}.
\adsurl{2005ApJ...628L.163W}.
\end{barticle}
\endbibitem

\bibitem[\protect\citeauthoryear{{Wuelser}
  \textit{et~al.}}{2004}]{Wuelser:2004}
\begin{bchapter}
\bauthor{\bsnm{{Wuelser}}, \binits{J.-P.}},
\bauthor{\bsnm{{Lemen}}, \binits{J.R.}},
\bauthor{\bsnm{{Tarbell}}, \binits{T.D.}},
\bauthor{\bsnm{{Wolfson}}, \binits{C.J.}},
\bauthor{\bsnm{{Cannon}}, \binits{J.C.}},
\bauthor{\bsnm{{Carpenter}}, \binits{B.A.}}:
\byear{2004},
\bctitle{{EUVI: the STEREO-SECCHI extreme ultraviolet imager}}.
In: \beditor{\bsnm{{Fineschi}}, \binits{S.}},
\beditor{\bsnm{{Gummin}}, \binits{M.A.}} (eds.)
\bbtitle{Telescopes and Instrumentation for Solar Astrophysics},
\bsertitle{\procspie}
\bseriesno{5171},
\bfpage{111}.
\doiurl{10.1117/12.506877}.
\adsurl{2004SPIE.5171..111W}.
\end{bchapter}
\endbibitem

\end{thebibliography}


\end{article} 

\end{document}